\pdfoutput=1
\documentclass[acmlarge,screen]{acmart}

\AtBeginDocument{%
  }

\setcopyright{acmlicensed}
\copyrightyear{2026}
\acmYear{2026}
\acmDOI{XXXXXXX.XXXXXXX}

\acmJournal{JDS}
\acmVolume{37}
\acmNumber{4}
\acmArticle{111}
\acmMonth{8}

\usepackage{xcolor}
\usepackage{etoolbox}
\patchcmd{\orcid}{\unskip}{\ifhmode\unskip\fi}{}{}

\usepackage{amsmath,amsfonts,amsthm}
\usepackage{array}
\usepackage{graphicx}
\usepackage{url}

\usepackage{booktabs}
\usepackage{multirow}
\usepackage[T1]{fontenc}
\usepackage{tikz}
\usepackage{forest}
\usetikzlibrary{graphs}
\usetikzlibrary{shapes.geometric, arrows.meta, positioning}
\usetikzlibrary{decorations.pathmorphing, shadows, 3d}
\usepackage{microtype}
\usepackage{makecell}
\usepackage{threeparttable}
\usepackage{enumitem}
\usepackage{wrapfig}
\usepackage{hyperref}
\hypersetup{
  colorlinks=true,
  linkcolor=blue,
  citecolor=blue,
  urlcolor=blue
}

\definecolor{hidden-draw}{RGB}{125,174,224}
\definecolor{hidden-orange}{RGB}{125,174,224}

\renewcommand{\citet}{\cite}

\begin{document}

\title{Time Series Analysis in Frequency Domain: A Survey of Open Challenges, Opportunities and Benchmarks}

\author{Regina (Qianru) Zhang}
\email{zqrhku@connect.hku.hk}
\orcid{0000-0002-5843-6187}
\affiliation{%
  \institution{The University of Hong Kong}
  \city{Hong Kong}
  \country{China}
}

\author{Yuting Sun}
\email{yuting.sun@uqconnect.edu.au}
\orcid{0000-0003-4482-4266}
\affiliation{%
  \institution{The University of Queensland}
  \city{Brisbane}
  \country{Australia}
}

\author{Honggang Wen}
\email{whgtytyg@gmail.com}
\orcid{0009-0006-8691-8569}
\affiliation{%
  \institution{The University of Hong Kong}
  \city{Hong Kong}
  \country{China}
}

\author{Peng Yang}
\email{stuyangpeng@gmail.com}
\orcid{0009-0008-5233-2986}
\affiliation{%
  \institution{The University of Hong Kong}
  \city{Hong Kong}
  \country{China}
}

\author{Xinzhu Li}
\email{lixinzhu544@gmail.com}
\orcid{0009-0007-1043-0863}
\affiliation{%
  \institution{The University of Hong Kong}
  \city{Hong Kong}
  \country{China}
}

\author{Ming Li}
\email{mingli@zjnu.edu.cn}
\orcid{0000-0002-1218-2804}
\affiliation{%
  \institution{Zhejiang Normal University}
  \city{Zhejiang}
  \country{China}
}

\author{Kwok-Yan Lam}
\email{kwokyan.lam@ntu.edu.sg}
\orcid{0000-0001-7479-7970}
\affiliation{%
  \institution{Nanyang Technological University}
  \city{Singapore}
  \country{Singapore}
}

\author{Pietro Liò\textsuperscript{*}}
\email{pl219@cam.ac.uk}
\orcid{0000-0002-0540-5053}
\affiliation{%
  \institution{The University of Cambridge}
  \city{Cambridge}
  \country{UK}
}

\author{Siu-Ming Yiu\textsuperscript{*}}
\email{smyiu@cs.hku.hk}
\orcid{0000-0002-3975-8500}
\affiliation{%
  \institution{The University of Hong Kong}
  \city{Hong Kong}
  \country{China}
}

\author{Hongzhi Yin\textsuperscript{*}}
\email{db.hongzhi@gmail.com}
\orcid{0000-0003-1395-261X}
\affiliation{%
  \institution{The University of Queensland}
  \city{Brisbane}
  \country{Australia}
}

\thanks{* Corresponding authors.}

\renewcommand{\shortauthors}{Zhang et al.}

\begin{abstract}
  Frequency-domain analysis has emerged as a powerful paradigm for time series analysis, offering unique advantages over traditional time-domain approaches while introducing new theoretical and practical challenges. This survey provides a comprehensive examination of spectral methods from classical Fourier analysis to modern neural operators, systematically summarizing three open challenges in current research: (1) causal structure preservation during spectral transformations, (2) uncertainty quantification in learned frequency representations, and (3) topology-aware analysis for non-Euclidean data structures. Through rigorous reviewing of over 100 studies, we develop a unified taxonomy that bridges conventional spectral techniques with cutting-edge machine learning approaches, while establishing standardized benchmarks for performance evaluation. Our work identifies key knowledge gaps in the field, particularly in geometric deep learning and quantum-enhanced spectral analysis. The survey offers practitioners a systematic framework for method selection and implementation, while charting promising directions for future research in this rapidly evolving domain.
\end{abstract}



\received{24 Sept. 2025}
\received[revised]{XX XX 2026}
\received[accepted]{XX XX 2026}

\maketitle

\section{Introduction}
\label{sec:intro}

The temporal dimension of data represents both an indispensable analytical resource and a formidable intellectual challenge across modern scientific inquiry. Time series analysis~\cite{mahalakshmi2016survey,sapankevych2009time,han2019review,zhang2026hmamba,zhang2026m} has emerged as a cornerstone of data-driven decision-making across a wide range of domains, including finance~\cite{zhang2025autohformer,han2019review,zhang2025fldmamba}, urban infrastructure management~\cite{zhang2025efficient,zhang2023automated,zhang2023spatial,zhang2025fldmamba,zhang2025billiards}, and climate system modeling~\cite{bracco2025machine}. This disciplinary breadth underscores the critical importance of developing robust analytical frameworks capable of extracting meaningful patterns from chronological observations while respecting their inherent nonstationarity and complex dependency structures.

\begin{figure*}[htb!]
\vspace{-0.15in}
\centering
\includegraphics[width=0.95\linewidth]{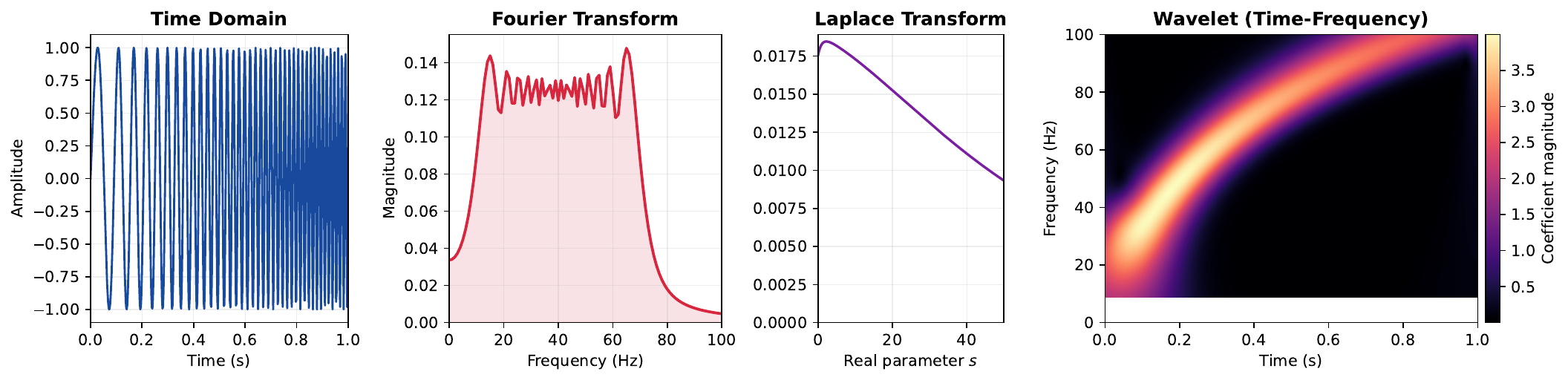}
\vspace{-0.15in}
\caption{Comparative visualization of spectral decomposition techniques demonstrating the fundamental trade-offs between temporal and frequency resolution}
\label{fig:intro_fig}
\vspace{-0.2in}
\end{figure*}

Conventional time-domain methodologies~\cite{wu2023timesnet,li2023revisiting,das2023long,wu2021autoformer,huang2024long,liu2023itransformer} remain effective and computationally attractive, but their assumptions and inductive biases can be restrictive for some nonstationary, multiscale, or long-range processes. For example, standard ARIMA models generally require an appropriate treatment of nonstationarity, while classical exponential-smoothing formulations represent level, trend, and seasonality rather than arbitrary cross-series spatial dependence. These limitations motivate frequency-domain representations as a complementary modeling perspective; they do not imply that time-domain methods are intrinsically inadequate.

Frequency-domain analysis offers a complementary representation~\cite{zhou2022fedformer}, transforming direct temporal modeling into the analysis of spectral components. Fourier decomposition has developed into a broad analytical tradition that includes wavelet theory, joint time--frequency analysis, and spectral graph methods. This breadth also complicates method selection. A global Fourier transform provides frequency information but does not localize transient events in time, whereas wavelet methods trade time and frequency resolution and require choices of basis and scale. Extensions are needed when observations lie on graphs or manifolds, because the relevant spectral basis then depends on the underlying geometry.

Figure~\ref{fig:intro_fig} provides a compelling visual metaphor for these analytical trade-offs. The Fourier spectrum's pristine frequency resolution stands in stark contrast to its temporal smearing of transient events, while wavelet transforms achieve superior time-frequency localization at the cost of basis function selection complexity. This dichotomy exemplifies the central tension in spectral methods: the irreducible compromise between frequency resolution and temporal precision. Modern applications increasingly demand techniques that transcend these classical limitations, particularly when analyzing signals exhibiting nonlinear manifold constraints, multiscale phenomena with abrupt regime shifts, or irregular sampling patterns characteristic of medical monitoring scenarios.

\begin{table*}
\centering
\vspace{-0.1in}
\caption{Time Series Surveys: Evolution and Coverage}
\vspace{-0.15in}
\label{tab:survey_comparison}
\resizebox{0.99\textwidth}{!}{
\setlength{\tabcolsep}{4mm}{
\begin{tabular}{clccl}
\toprule
\textbf{Survey} & \textbf{Focus} & \textbf{Spectral Coverage} & \textbf{Year} & \textbf{Key Limitation} \\
\midrule
\cite{esling2012time} & Similarity measures & Limited/secondary & 2012 & No modern methods \\
\cite{sapankevych2009time} & Classical TS & Limited/secondary   & 2009 & Time-domain focus \\
\cite{lim2021time} & Deep learning & Limited/secondary & 2021 & Minimal spectral analysis \\
\cite{guo2022wavelet} & Wavelet theory and applications & Wavelets; wavelet--NN integration & 2022 & Wavelet-specific scope \\
\cite{kovachki2023neural} & Neural operators & Neural operators, including spectral operators & 2023 & No benchmarks \\
\cite{liang2024foundation} & Foundation models & Limited/secondary & 2024 & Weak on interpretability \\
\cite{wen2022transformers} & Transformers & Limited/secondary & 2023 & Minimal spectral analysis \\
\cite{liu2024timemmd} & Multimodal TS & Limited/secondary & 2024 & Time-domain focus \\
\hline
\textbf{Ours} & \textbf{Unified spectrum} & \textbf{Fourier; wavelets; time--frequency; neural spectral; hybrid} & \textbf{2026} & \textbf{Broad synthesis; no unified reruns} \\
\bottomrule
\end{tabular}}}
\vspace{-0.2in}
\begin{minipage}{0.95\linewidth}
\footnotesize
\vspace{2pt}
\textit{Note.} The former scalar ``$x/5$'' depth score has been removed because a single number obscures which topics a survey actually treats and can conflate publication date with quality. The table now reports conservative, topic-level spectral coverage. ``Limited/secondary'' means that spectral analysis is not a principal organizing topic of the cited survey. The broad methodological landmarks used to organize this survey are global harmonic analysis (Fourier), localized multiresolution analysis (wavelets), classical joint time--frequency representations such as STFT and Wigner--Ville methods, learned spectral mappings (neural operators), and integration with physical or causal structure. These are organizing categories rather than a claim that each began in a single decade or forms a strictly linear historical stage.
\end{minipage}
\end{table*}

{\textbf{Related Surveys and Motivations of This Work}} The historical trajectory of frequency-domain methodologies, extending from Fourier's foundational harmonic analysis to contemporary neural spectral operators (Table~\ref{tab:survey_comparison}), has yielded a vast yet disorganized corpus of research. While existing surveys~\cite{liang2024foundation,zhang2024large,wang2024deep,wen2022transformers} have examined discrete segments of this continuum, they collectively fail to deliver a deep analysis of three pivotal contemporary challenges. First is \textbf{the preservation of causal structure during spectral transformations}, which addresses the critical need to ensure models learn logical cause-and-effect relationships, preventing flawed forecasts where an effect precedes its cause. Second, they lack guidance on \textbf{rigorous uncertainty quantification in learned frequency representations}, a key step for establishing model reliability in high-stakes domains like medical diagnostics, where a model's confidence is as important as its prediction. Finally, they overlook the need for \textbf{topology-adaptive spectral analysis for non-Euclidean data}, an essential innovation for applying these powerful methods to complex, real-world networks like brain connectomes or supply chains whose irregular structures defy traditional techniques. This critical oversight has created a methodological vacuum, leaving practitioners without guidance for algorithm selection, implementation, or unified taxonomies. Our work bridges this gap by delivering not only comprehensive categorization but also incisive analysis of these open challenges, paired with viable solution pathways grounded in cutting-edge developments.

\textbf{Paper selection} This survey focuses on more than 100 works from 2020--2025, supplemented by earlier foundational references. The corpus includes peer-reviewed conference and journal papers as well as clearly identified preprints, theses, technical reports, and official dataset records. As illustrated in Fig.~\ref{fig:taxonomy}, these works are organized into two categories: (1) methodological innovations, including spectral transforms, neural operators, and hybrid architectures; and (2) application domains, including time series forecasting, time series anomaly detection, and related tasks.

Our survey makes three substantive contributions to advancing the field:
\begin{itemize}
    \item We develop a unified taxonomic framework that encompasses both classical and modern frequency-domain techniques and systematically organizes them according to their mathematical foundations, computational characteristics, and application domains.

    \item We provide a structured synthesis of results reported for representative methods on real-world datasets, while explicitly documenting that differences in splits, preprocessing, tuning, and evaluation protocols preclude a controlled ranking.

    \item We identify emerging research frontiers at the intersection of spectral analysis and geometric deep learning, with particular emphasis on quantum-enhanced time-frequency analysis and topology-preserving spectral methods.
\end{itemize}

Through this comprehensive examination of established and cutting-edge approaches, our work establishes new evaluation standards while providing practitioners with principled guidance for navigating the increasingly sophisticated landscape of temporal data analysis. The comprehensive taxonomy, clear pipeline, and domain-specific insights position this survey as an essential reference for researchers confronting the open challenges and opportunities in frequency-domain time series analysis.

\section{Survey Architecture and Pipeline}\label{sec:architecture}
\begin{figure*}
    \centering
    \includegraphics[width=0.95\linewidth]{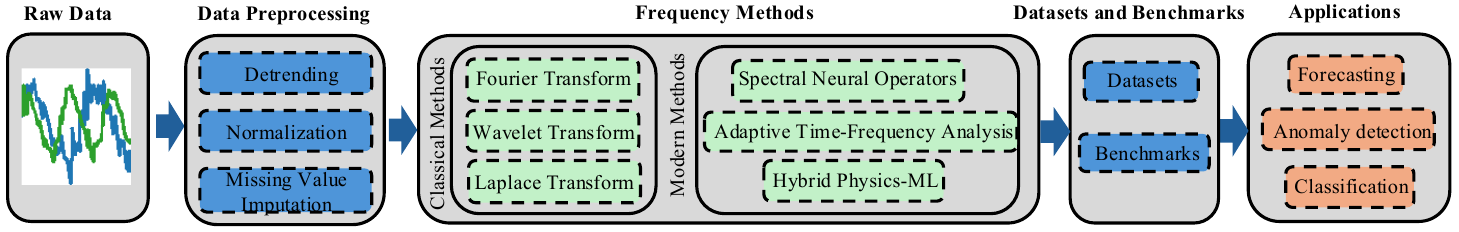}
    \vspace{-0.1in}
    \caption{Pipeline for frequency-domain time-series analysis. A dataset is a corpus that may contain raw or prepared observations. ``Raw data'' denotes unprocessed readings, ``preprocessed data'' denotes signal-ready inputs after conditioning (Section~\ref{sec:data_pre}), and ``benchmarking'' denotes evaluation of methods on standardized dataset splits (Section~\ref{sec:benchmark}). Thus, datasets are inputs to the workflow rather than a downstream processing stage.}
    \vspace{-0.1in}
    \label{fig:frame_pipeline}
\end{figure*}

This comprehensive survey is organized as follows: Section~\ref{sec:prob} formally establishes the problem space of frequency-based time series analysis.  Section~\ref{sec:pipeline} presents our systematic analysis framework, visually supported by Figure~\ref{fig:frame_pipeline}, while Section~\ref{sec:data_pre} details essential preprocessing techniques. Section~\ref{sec:challenges} presents open challenges of time series analysis. The methodological core comprises: (1) Section~\ref{sec:class}, reviewing classical spectral transformations (Fourier, Wavelet, and Laplace); and (2) Section~\ref{sec:current}, examining modern advances like spectral neural operators. Rigorous empirical evaluation across standard benchmarks follows in Section~\ref{sec:benchmark}. We then analyze domain-specific applications in Section~\ref{sec:application}, identify current limitations in detail in Section~\ref{sec:challenge_details}, and outline emerging directions in Section~\ref{sec:future}. The survey concludes with synthesized insights in Section~\ref{sec:conclu}. For structural clarity, Figure~\ref{fig:taxonomy} presents our hierarchical taxonomy of time series analysis methodologies.

\subsection{Problem Definition}\label{sec:prob}
Let $\mathcal{X} = (x_1, \dots, x_L) \in \mathbb{R}^{L \times V}$ denote an input time series with $L$ time steps and $V$ variables, where each observation $x_i \in \mathbb{R}^V$ is a $V$-dimensional feature vector. For different analytical tasks, we define:

\begin{itemize}[leftmargin=2em]
    \item \textbf{Forecasting}: Predict future values $\mathcal{Y}^{f} = (x_{L+1}, \dots, x_{L+H}) \in \mathbb{R}^{H \times V}$ with horizon $H$

    \item \textbf{Anomaly Detection}: Identify anomalous points $\mathcal{Y}^{a} = (y^{a}_1, \dots, y^{a}_L) \in \{0,1\}^L$ where $y^{a}_t=1$ marks anomalies

    \item \textbf{Classification}: Assign a class label $y^{c} \in \mathcal{Y}_c$, where $\mathcal{Y}_c=\{1, \dots, K\}$ contains $K$ predefined categories
\end{itemize}

\noindent \textbf{Frequency Transform.} To enhance pattern recognition across these tasks, we employ a generic spectral operator $\mathcal{T}(\cdot)$, which produces a shared frequency-domain representation $\mathbf{Z}$:

\begin{equation}
    \label{equation1}
    \mathbf{Z} = \mathcal{T}(\mathcal{X}) = \{\mathcal{T}_v(x_{:,v})\}_{v=1}^V
\end{equation}

where $\mathcal{F}$ represents traditional transforms
 like Fourier/Wavelet/Laplace transforms, and modern transforms like spectral neural operators. Here, $\mathcal{T}_v$ denotes the selected transform applied to variable $v$; it is not itself a task predictor. Distinct task-specific models then map $\mathbf{Z}$ to the respective outputs: $\mathcal{Y}^{f} = \mathcal{M}_f(\mathbf{Z})$ for forecasting, $\mathcal{Y}^{a} = \mathcal{M}_a(\mathbf{Z})$ for anomaly detection, and $y^{c} = \mathcal{M}_c(\mathbf{Z})$ for classification. This transformation enables:

\begin{itemize}[leftmargin=2em]
    \item \textit{Forecasting}: Explicit periodicity modeling through dominant frequency components

    \item \textit{Anomaly Detection}: Deviation identification in spectral power distributions

    \item \textit{Classification}: Discriminative feature extraction via frequency signatures
\end{itemize}

The key challenge lies in jointly optimizing temporal and spectral representations for multi-task learning while maintaining computational efficiency.

\subsection{Pipeline of Frequency Methods for Time Series Analysis} \label{sec:pipeline}
Our frequency-domain analysis pipeline (Figure~\ref{fig:frame_pipeline}) establishes a systematic framework comprising four critical stages: \textit{data preprocessing} (Section~\ref{sec:data_pre}) for signal conditioning, \textit{frequency transformation} through both traditional spectral methods (Fourier/Wavelet transforms; Section~\ref{sec:class}) and modern neural approaches (Section~\ref{sec:current}), rigorous \textit{evaluation} on standardized benchmark datasets (Section~\ref{sec:benchmark}), and practical \textit{deployment} in domains like healthcare monitoring and financial forecasting (Section~\ref{sec:application}). A dataset may contain raw or prepared observations; benchmarking is the evaluation activity that aggregates reported results across multiple methods and standardized dataset splits. This end-to-end workflow ensures methodological consistency from raw data ingestion ($x_t \in \mathbb{R}^V$) to task-specific outputs ($\mathcal{Y}^{f}, \mathcal{Y}^{a}, y^{c}$), where each stage's design choices directly influence downstream performance metrics. The subsequent sections detail technical implementations and specific descriptions at each phase.

\subsection{Open Challenges for Time Series Analysis}
\label{sec:challenges}

\begin{figure}[!htbp]
    \centering
    \includegraphics[width=0.80\linewidth, height=0.27\linewidth]{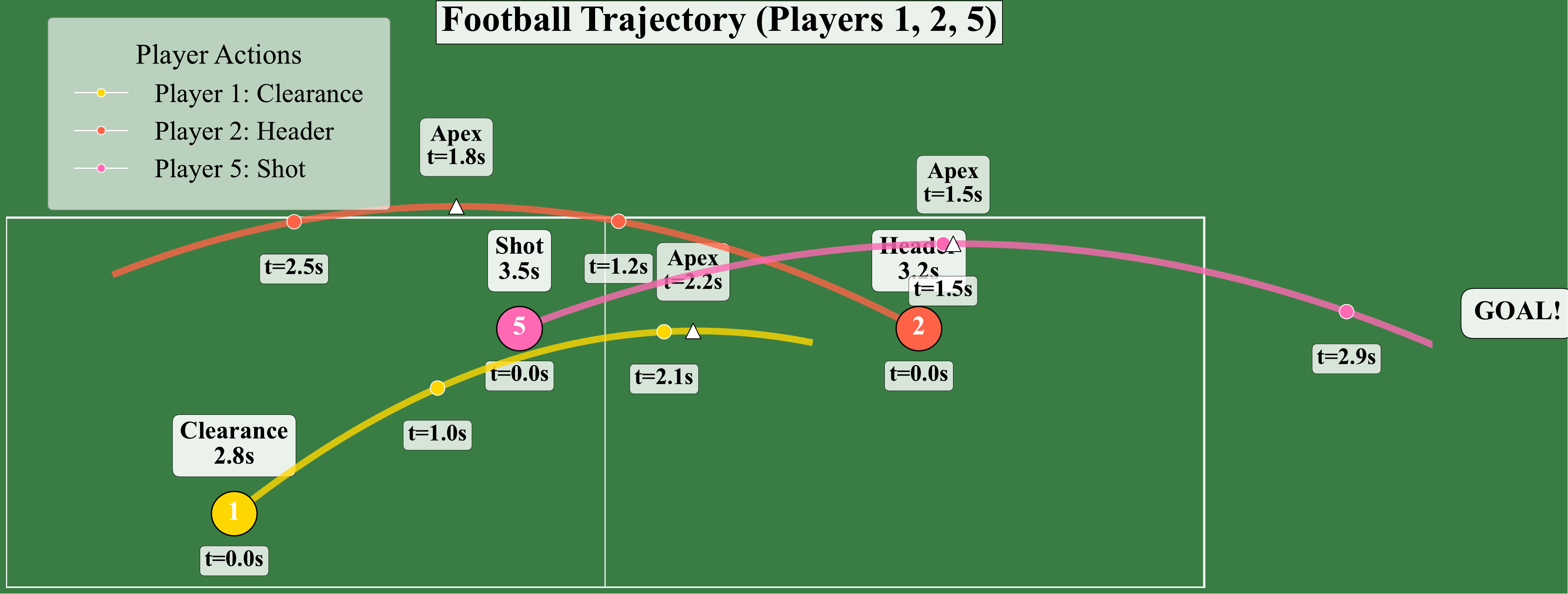}
    \vspace{-0.1in}
    \caption{Nonstationary dynamics in football trajectory time series. The piecewise parabolic segments (Player 1: clearance, Player 2: header, Player 5: shot) demonstrate abrupt parameter shifts where velocity $v(t)$ and launch angle $\theta(t)$ change discontinuously at each player interaction ($t_1,t_2,t_3$). Traditional temporal models fail to capture these regime shifts, as the system dynamics $\frac{d\mathbf{x}}{dt} = f_t(\mathbf{x})$ vary unpredictably with player interventions.}
    \vspace{-0.1in}
    \label{fig:foot_traj}
\end{figure}

Time series analysis in the \textit{temporal domain} faces inherent limitations, including error accumulation in long sequences, local receptive field constraints, and nonstationary dynamics, which are partially mitigated by techniques like dilated convolutions and attention mechanisms, shown in Table~\ref{tab:tradeoffs}. For example, we present nonstationary dynamics in football trajectory time series in Figure~\ref{fig:foot_traj}. The trajectories exemplify three key aspects of nonstationary time series: \textit{1) Parameter discontinuity}: Ball dynamics reset at each player contact (e.g., $v_{t_2^+} \neq v_{t_2^-}$); \textit{2) Context-dependent transitions}: Header-induced trajectory change ($\Delta\theta > 90^\circ$) differs fundamentally from shot dynamics; \textit{3) Unobservable switches}: The exact moment of player contact (dashed circles) is often unknown to temporal models.
This illustrates why adaptive RNNs and change-point detection remain incomplete solutions - they cannot anticipate future player interactions that fundamentally alter the system's governing equations. This example can also be viewed through a spatio-temporal causal lens: player actions can be represented as intervention or event variables that alter dynamical parameters ($v$, $\theta$). Whether an action is exogenous depends on the assumed causal graph; under either interpretation, causal structure recovery and intervention modeling complement frequency-domain analysis~\cite{azad2024vision,besserve2022cause}.

The above challenges in the temporal domain often motivate complementary analysis in the \textit{frequency domain} (Section~\ref{sec:challenge_details}). Spectral methods offer global signal representations, but curved manifolds $\mathcal{M}$, non-uniform sampling, and nonlinear embeddings introduce additional difficulties. In particular, (1) manifold curvature distorts standard Fourier harmonics, requiring basis functions that respect the metric tensor $g_{ij}$ (the components of the local inner product that determines distances and angles on $\mathcal{M}$); (2) non-uniform sampling violates the regular-grid assumption of the standard FFT and can create biased spectral estimates; and (3) a nonlinear embedding $\phi:\mathcal{M}\to\mathbb{R}^d$ can make learned frequencies sensitive to perturbations. We denote this sensitivity conceptually by $\delta_{\omega_k}$, the change in the $k$-th estimated frequency under a specified perturbation; its numerical value depends on the estimator, perturbation model, and embedding and is therefore not assigned a universal propagation formula here. Existing work addresses several distinct notions of causality. Spectral-independence inference~\cite{besserve2022cause} and FFT-based structure recovery~\cite{veedu2023fftcausal} provide identification results for linear dynamical systems. By contrast, the causality-preserving Fourier method of~\cite{knight2019causality} preserves the temporal direction of a specified atmospheric-wave solver, and NeuSA~\cite{bizzi2025neusa} enforces temporal ordering in an initial-value PDE architecture; neither is a general causal-discovery result. General causal identification under nonlinear dynamics, latent confounding, irregular sampling, or non-Euclidean state spaces remains open.

\begin{table}[h]
    \centering
    \vspace{-0.1in}
    \caption{Open Challenges of Temporal Series Analysis in Time and Frequency Domain}
    \vspace{-0.15in}
    \label{tab:tradeoffs}
    \resizebox{0.99\textwidth}{!}{\begin{tabular}{@{}lll@{}}
    \toprule
    \textbf{Domain} & \textbf{Open Challenges} & \textbf{Current Solutions} \\
    \midrule
    \multirow{4}{*}{Temporal}
    & 1. Error accumulation in long sequences & Dilated convolutions \\
    & 2. Local receptive fields & Attention mechanisms \\
    & 3. Nonstationary dynamics & Adaptive RNNs \\
    & 4. Unpredictable regime shifts & Change-point detection \\
    \addlinespace
        \multirow{7}{*}{Frequency}
    & 1. Fixed time-frequency resolution & Wavelet packets \\
    & 2. Spectral leakage & Multitaper methods \\
    & 3. Cross-term interference & Wigner-Ville decomposition \\
    & 4. Basis mismatch & Dictionary learning \\
    & \emph{Still open in general:} 5. Causal identification under nonlinear or latent-variable settings & Partial identification results for linear systems~\cite{besserve2022cause,veedu2023fftcausal} \\
    & \emph{Still Persist:} 6. Topological obstructions & -- \\
    & \emph{Still Persist:} 7. Quantifying uncertainty& -- \\
    \bottomrule
    \end{tabular}}
    \vspace{-0.2in}
\end{table}

\begin{figure}[h]
\vspace{-0.15in}
\centering
\begin{tikzpicture}[scale=1.8, >=Stealth,
    declare function={
        f(\x,\y) = -0.3*(\x-2)^2 - 0.2*\y^2 + 0.03*\x*\y;
    }]

\foreach \i in {0,0.1,...,2.5} {
    \draw[blue!30, thin, shift={(0,0,{f(\i,0)})}] (\i,0) arc (0:-180:{\i} and {\i/3});
}
\fill[blue!15, opacity=0.8, domain=0:2.5, samples=30]
    plot (\x,0,{f(\x,0)}) --
    plot[domain=2.5:0] (\x,0,{f(\x,0)-0.3}) -- cycle;
\fill[top color=blue!20, bottom color=blue!5, opacity=0.7, domain=0:2.5, samples=30]
    plot (\x,0,{f(\x,0)}) --
    plot[domain=2.5:0] (\x,1,{f(\x,1)}) -- cycle;

\draw[red, ultra thick, ->]
    plot[domain=0.5:3.5, samples=30, smooth]
    (\x-0.5,{0.2*sin(\x*120)}, {f(\x-0.5,{0.2*sin(\x*120)})+0.05});

\foreach \t/\lab in {0.5/$x_1$,1.7/$x_2$,2.6/$x_3$,3.2/$x_4$} {
    \fill[red, shift={({\t-0.5},{0.2*sin(\t*120)},{f({\t-0.5},{0.2*sin(\t*120)})+0.05})}]
        circle (1pt) node[white, fill=red, circle, inner sep=1pt, outer sep=2pt] {\lab};
}

\begin{scope}[shift={({1.2},{0.2*sin(1.7*120)},{f(1.2,{0.2*sin(1.7*120)})+0.05})}, rotate around={15:(0,0,1)}]
    \fill[gray!20, opacity=0.9] (0,0) -- (1,0.5) -- (1,-0.5) -- cycle;
    \draw[->, thick] (0,0) -- (0.7,0.35) node[right, black] {$T_{x}\mathcal{M}$};
    \draw[->, thick] (0,0) -- (0.7,-0.35);
\end{scope}

\begin{scope}[canvas is yz plane at x=5]
    \fill[green!15, opacity=0.8] (0,0) circle (0.8 and 0.4);
    \draw[red, thick] plot[domain=0:360, samples=60, smooth]
        ({0.3*cos(\x) + 0.1*cos(3*\x)}, {0.15*sin(\x) + 0.05*sin(2*\x)});
    \foreach \t/\lab in {30/$z_1$,180/$z_3$} {
        \fill[red] ({0.3*cos(\t) + 0.1*cos(3*\t)}, {0.15*sin(\t) + 0.05*sin(2*\t)})
            circle (1pt) node[below, black] {\lab};
    }
\end{scope}
\draw[->, very thick] (3.5,0,0) -- (4.5,0,0) node[midway,above] {$\phi$};

\shade[ball color=red!80, opacity=0.9]
    ({1.2},{0.2*sin(1.7*120)},{f(1.2,{0.2*sin(1.7*120)})+0.1}) circle (0.8mm);

\node[blue!70!black, font=\small\bfseries] at (1.5,1.5,{f(1.5,1.5)+0.3}) {$\mathcal{M}$};
\node[anchor=west, font=\scriptsize] at (5.2,0.8,0) {Latent Space};

\end{tikzpicture}
    \vspace{-0.1in}
    \caption{Three-dimensional visualization of time series data evolving on a curved Riemannian manifold $\mathcal{M}$. The red trajectory represents the temporal dynamics $\{x_t\}$, with the tangent space $T_x\mathcal{M}$ showing local linearization. The green ellipse depicts the nonlinear embedding $\phi:\mathcal{M}\to\mathbb{R}^d$ preserving temporal relationships as $\{z_t\}$.}
    \label{fig:3d_riemannian}
    \vspace{-0.2in}
\end{figure}

\subsection{Data Preprocessing}\label{sec:data_pre}
This section introduces three fundamental preprocessing techniques for time series analysis: \textbf{detrending}, \textbf{normalization}, and \textbf{missing value imputation}. Each method addresses distinct challenges in raw temporal data preparation. Details of each mechanism are shown as follows:

\textbf{Detrending}. Detrending is a critical preprocessing technique that eliminates long-term trends from time series data to enhance analytical accuracy. By removing non-stationary components (e.g., linear, polynomial, or exponential trends), it reveals underlying patterns and improves forecasting model performance. Common approaches include: (1) \textit{differencing} ($\nabla x_t = x_t - x_{t-1}$) for removing stochastic trends, (2) \textit{regression detrending} using ordinary least squares (OLS), and (3) \textit{filter-based methods} like the Hodrick-Prescott (HP) filter. Detrending is particularly vital for economic indicators, climate data, and sensor signals where trends may obscure true relationships. However, improper application can lead to over-differencing or signal distortion. The choice between methods depends on trend characteristics, with unit root tests (e.g., Augmented Dickey-Fuller) guiding selection. Effective detrending stabilizes variance and enables reliable application of ARIMA, spectral analysis, and machine learning techniques.

\noindent \textbf{Normalization}. Normalization transforms time series data to a common scale while preserving temporal relationships, essential for multi-variate analysis and machine learning. Standard techniques include: (1) \textit{Z-score normalization} ($z_t = \frac{x_t - \mu}{\sigma}$) for Gaussian-distributed data, (2) \textit{Min-Max scaling} ($x'{t} = \frac{x_t - x{\min}}{x_{\max} - x_{\min}}$) bounding values to $[0,1]$, and (3) \textit{Decimal scaling} for preserving sign. Normalization mitigates feature dominance in models like RNNs and transformers, improves gradient descent efficiency, and enables meaningful distance comparisons (e.g., DTW). Critical applications span sensor fusion, financial forecasting, and biomedical signal processing. However, normalizing non-stationary series requires windowed adaptation or online scaling. The choice depends on data distribution, outlier sensitivity, and downstream tasks, with robust scaling (e.g., using median/IQR) preferred for noisy real-world datasets.

\noindent \textbf{Missing Value Imputation}. Missing data imputation is critical for maintaining temporal integrity in time series analysis. Common approaches include: (1) \textit{Forward/Backward Filling} for short gaps, (2) \textit{Linear Interpolation} ($x_t = x_{t-1} + \frac{t - t_{t-1}}{t_{t+1} - t_{t-1}}(x_{t+1} - x_{t-1})$) for short gaps when change between the two bracketing observations is approximately linear, (3) \textit{Seasonal Imputation} using moving windows for periodic data, and (4) \textit{Model-Based Methods} (ARIMA, neural networks) for complex patterns. The choice depends on missingness mechanism (MCAR, MAR, MNAR), gap length, and data characteristics. While simple methods work for small gaps ($<$5\% missing), large gaps may require advanced techniques like multiple imputation or matrix factorization. Careful validation via artificial masking is essential to assess imputation quality without distorting autocorrelation structure or introducing bias in downstream forecasting tasks.

\tikzstyle{my-box}=[
    rectangle,
    draw=hidden-draw,
    rounded corners,
    align=left,
    text opacity=1,
    minimum height=1.3em,
    minimum width=5em,
    inner sep=2pt,
    fill opacity=.8,
    line width=0.8pt,
]
\tikzstyle{leaf-head}=[my-box, minimum height=1.5em,
    draw=hidden-orange, text=black, font=\normalsize,
    inner xsep=2pt,
    inner ysep=4pt,
    line width=0.8pt,
]
\tikzstyle{leaf-task}=[my-box, minimum height=2.5em,
    draw=hidden-orange, text=black, font=\normalsize,
    inner xsep=2pt,
    inner ysep=4pt,
    line width=0.8pt,
]
\tikzstyle{modelnode-task1}=[my-box, minimum height=1.5em,
    draw=hidden-orange, fill=hidden-draw, text=black, font=\normalsize,
    inner xsep=2pt,
    inner ysep=4pt,
    line width=0.8pt,
]
\tikzstyle{leaf-task10}=[my-box, minimum height=1.0em,
    draw=hidden-orange, text=black, font=\normalsize,
    inner xsep=2pt,
    inner ysep=4pt,
    line width=0.6pt,
]
\tikzstyle{leaf-paradigms}=[my-box, minimum height=2.5em,
    draw=hidden-orange, text=black, font=\normalsize,
    inner xsep=2pt,
    inner ysep=4pt,
    line width=0.8pt,
]
\begin{figure*}[htbp]
    \centering
    \resizebox{0.99\textwidth}{!}{
        \begin{forest}
            for tree={
                grow=east,
                reversed=true,
                anchor=base west,
                parent anchor=east,
                child anchor=west,
                base=left,
                font=\normalsize,
                rectangle,
                draw=hidden-draw,
                rounded corners,
                align=left,
                minimum width=1em,
                edge+={darkgray, line width=1pt},
                s sep=3pt,
                inner xsep=0pt,
                inner ysep=3pt,
                line width=0.8pt,
                ver/.style={rotate=90, child anchor=north, parent anchor=south, anchor=center},
                edge path={
                    \noexpand\path [draw, \forestoption{edge}]
                    (!u.parent anchor) -- ++(5pt,0) |- (.child anchor)\forestoption{edge label};
                },
            },
            [Frequency Transforms in Time Series Analysis,leaf-head,ver
                [
                     Frequency Transform \\ Methods (\textbf{Sec.~\ref{sec:class}} and\\ \textbf{Sec.~\ref{sec:current}}),leaf-task,text width=9em
                    [
                        Classical Methods, leaf-task10, text width=8em
                        [
                            Fourier Transform, leaf-task10, text width=14.5em
                            [NFT~\cite{koren2024interpretable}{, }Fournet~\cite{du2023fourier}{, }Pastnet~\cite{wu2024pastnet}{, }LPR~\\\cite{chen2024laplacian}{, }STFT~\cite{yao2019stfnets}{, }FrFT~\cite{kocc2022fractional}{, }SGFT~\cite{defferrard2016convolutional}{, }\\WGT~\cite{xu2019graph},modelnode-task1, text width=20em]
                        ]
                        [
                            Wavelet Transform, leaf-task10, text width=14.5em
                            [LGT~\cite{cosentino2020learnable}{, }WaveForM~\cite{yang2023waveform}{, }FEDformer~\cite{zhou2022fedformer}\\DWT~\cite{wen2021robustperiod}{, }WaveRoRA~\cite{liang2024waverora}{, }WFTNet~\cite{liu2024wftnet} , modelnode-task1, text width=20em]
                        ]
                        [
                            Laplace Transform, leaf-task10, text width=14.5em
                            [Ambhika \textit{et al.}~\cite{ambhika2024time}{, }Chen \textit{et al.}~\cite{chen2024laplacian}{, }\\Shu et al.~\cite{shu2024low}
                            , modelnode-task1, text width=20em]
                        ]
                        [
                            Other Transforms, leaf-task10, text width=14.5em
                            [Cheng \textit{et al.}~\cite{cheng2014novel}{,}CEEMD~\cite{leung2021financial}{,} Ma \textit{et al.}~\cite{hao2015analysis}{,} \\EMD–SAE~\cite{yang2019hybrid}{,} HHT-XGB~\cite{dezhkam2023forecasting}{,}\\ARIMA-ANN~\cite{buyukcsahin2019improving}
                            , modelnode-task1, text width=20em]
                        ]
                    ]
                    [
                        Modern Methods, leaf-task10, text width=8em
                        [
                            Spectral Neural Operators, leaf-task10, text width=14.5em
                            [Neural Chaos~\cite{bahmani2025neural}{, }CoNO~\cite{tiwari2025cono}{, }Du \textit{et al.}~\cite{du2023neural}{, }\\ SHIFTING TIME~\cite{damashifting}{, }Micha{\l}owska \textit{et al.}~\cite{michalowska2024neural}{, }\\Oommen \textit{et al.}~\cite{oommen2025integrating}{, }Spectral-Refiner~\cite{cao2025spectral}{,}\\ Long \textit{et al.}~\cite{long2024time}{, }Wang \textit{et al.}~\cite{wang2024chaotic}{, }\\ Zhang \textit{et al.}~\cite{zhang2025fourier}{, }Qin \textit{et al.}~\cite{qin2024toward}, modelnode-task1, text width=20em]
                        ]
                        [
                            Adaptive Time-Frequency Analysis, leaf-task10, text width=14.5em
                            [Chao \textit{et al.}~\cite{chao2025self}{, }Huang \textit{et al.}~\cite{huang2024adaptive}{, }Leiber~\cite{leiber2024adaptive}{, }\\ Faith~\cite{li2025faith}{, }Liang \textit{et al.}~\cite{liang2025adaptive}{,
                            }Warion \textit{et al.}~\cite{warion2024class}{, }\\ Luo \textit{et al.}~\cite{luo2024time}{,
                            }Lu \textit{et al.}~\cite{lu2025filtering}{, }Xu \textit{et al.}~\cite{xu2025local}{,
                            }\\ ATFNet~\cite{ye2024atfnet}{, }FAN-TSF~\cite{zhang2025fan}, modelnode-task1, text width=20em]
                        ]
                        [
                            Hybrid Physics-ML, leaf-task10, text width=14.5em
                            [Abbasi \textit{et al.}~\cite{abbasi2024physics}{, }Qi \textit{et al.}~\cite{qi2025coupled}{, }CARD~\cite{wang2024card}{, }\\
                            Hsieh \textit{et al.}~\cite{hsieh2024hybrid}{, }Jiao \textit{et al.}~\cite{jiao2024hybrid}{, }
                            Lin \textit{et al.}~\cite{lin2024stress}{, }\\ Lin \textit{et al.}~\cite{lin2023systematic}{, }
                            Lin \textit{et al.}~\cite{lin2023sampling}{, }Liu \textit{et al.}~\cite{liu2022hybrid}{, }\\
                            Christopoulos~\cite{christopoulos2025towards}
                            , modelnode-task1, text width=20em]
                        ]
                    ]
                ]
                [
                    Frequency Transform \\ Application (\textbf{Sec.~\ref{sec:application}}), leaf-paradigms,text width=9em
                    [
                        Forecasting, leaf-task10, text width=8em
                        [
                            Frequency-enhanced Transformer, leaf-task10, text width=14.5em
                            [Zhang \textit{et al.}~\cite{zhang2022less}{, }Ismail \textit{et al.}~\cite{ismail2020inceptiontime}{, }TST~\cite{jin2022time}{, }\\ Liu \textit{et al.}~\cite{liu2022scinet}{, }Wu \textit{et al.}~\cite{wu2023timesnet}{, }Xu \textit{et al.}~\cite{xu2022anomaly}{, }\\Yi \textit{et al.}~\cite{yi2023frequency}{, } Nie \textit{et al.}~\cite{nie2023time}{, }Bai \textit{et al.}~\cite{bai2018empirical}{, }\\Bansal \textit{et al.}~\cite{bansal2021missing}{, }Karim \textit{et al.}~\cite{karim2019multivariate}{, }CRT~\cite{zhang2023self}{, }\\Wang \textit{et al.}~\cite{wang2023micn}{, }Msgnet~\cite{cai2024msgnet}{, }TSLANet~\cite{eldele2024tslanet}{, }\\Lang \textit{et al.}~\cite{lange2021fourier}{, }Park \textit{et al.}~\cite{park2021fast}{, }Frnet~\cite{zhang2024frnet}{, }\\FreTS~\cite{yi2023frequency}{, }TSLANet~\cite{eldele2024tslanet}{, }Informer~\cite{zhou2021informer}{, }\\Tran \textit{et al.}~\cite{tran2023fourier}{, }FEDAF~\cite{yang2024fedaf}{, }Yi \textit{et al.}~\cite{yi2023survey}{, }\\FEDformer~\cite{zhou2022fedformer}{, }Reformer~\cite{kitaev2020reformer}{, }Ma \textit{et al.}~\cite{ma2023long}{, }\\CARD~\cite{wang2024card}{, }Zhou \textit{et al.}~\cite{zhou2024fourier}{, }DEPTS~\cite{fan2022depts}{, } \\FiLM~\cite{zhou2022film}{, }FourierGNN~\cite{yi2023fouriergnn}{, }FITS~\cite{xu2024fits}{, }\\PFT~\cite{park2021fast}{, }FAN-TSF~\cite{zhang2025fan}{, }Fredformer~\cite{piao2024fredformer}{, }\\Cheng \textit{et al.}~\cite{cheng2023formertime}{, }FreEformer~\cite{yue2025freeformer}{, } \\Scaleformer~\cite{shabani2023scaleformer}{, }Autoformer~\cite{wu2021autoformer}{, }\\FEDformer~\cite{yang2024fedaf}{, }W-Transformer~\cite{sasal2022w}{, }\\ETSFormer~\cite{woo2022etsformer}\\,modelnode-task1, text width=20em]
                        ]
                        [
                            Multi-scale Decomposition, leaf-task10, text width=14.5em
                            [Vaswani \textit{et al.}~\cite{vaswani2017attention}{, }Zeng \textit{et al.}~\cite{zeng2023transformers}{, }\textit{etc }, modelnode-task1, text width=20em]
                        ]
                    ]
                    [
                       Anomaly detection, leaf-task10, text width=8em
                       [
                            Spectral Signature Analysis, leaf-task10, text width=14.5em
                            [Elsner \textit{et al.}~\cite{elsner2013singular}{, }Shi \textit{et al.}~\cite{shi2025characterizing}{, }\textit{etc }, modelnode-task1, text width=20em]
                        ]
                        [
                            Wavelet-based Change Detection, leaf-task10, text width=14.5em
                            [Ebrahim \textit{et al.}~\cite{ghaderpour2021survey}{, }Fonseca \textit{et al.}~\cite{fonseca2024time}{, }\textit{etc }, modelnode-task1, text width=20em]
                        ]
                    ]
                    [
                       Classification, leaf-task10, text width=8em
                       [
                            Discriminative Frequency Features, leaf-task10, text width=14.5em
                            [TSCMamba~\cite{ahamed2025tscmamba}{, }Fan \textit{et al.}~\cite{fan2025towards}{, }\textit{etc}, modelnode-task1, text width=20em]
                        ]
                        [
                            Cross-domain Representations, leaf-task10, text width=14.5em
                            [Cheng \textit{et al.}~\cite{cheng2025cross}{, }Hu \textit{et al.}~\cite{hu2025fusion}{, }\textit{etc}, modelnode-task1, text width=20em]
                        ]
                    ]
                ]
            ]
        \end{forest}
    }
    \caption{Taxonomy of frequency transform in time series analysis}
    \label{fig:taxonomy}
\end{figure*}

\section{Classical Methodological Advances}\label{sec:class}

\subsection{Classical Methods for Frequency Transform}\label{sec:classical-frequency-transform-approaches}

Frequency transforms are categorized into Fourier, wavelet, and Laplace transforms based on their formulations and applications. Furthermore, Table~\ref{tab:frequency_transform_methods} summarizes the representative frequency transform methods.

\begin{table*}[!ht]
    \centering
    \renewcommand\arraystretch{1.5}
    \caption{Summary of representative frequency transform methods in our framework.}
    \label{tab:frequency_transform_methods}
    \vspace{-0.4cm}
    \scriptsize
    \setlength{\tabcolsep}{6pt}
    \resizebox{\textwidth}{!}{
        \fontsize{12}{1.25\baselineskip}\selectfont
        \begin{tabular}{c|c|c|l}
        \toprule[1.2pt]
        \textbf{\makecell[c]{Frequency \\ Transform}} & \textbf{\makecell[c]{Categories and \\ Representative Methods}} & \textbf{Expression} & \textbf{Notes (Advantages \& Disadvantages)} \\
        \midrule[1.2pt]
        \multirow{6}{*}{\makecell[c]{\\ \textbf{Fourier} \\ \textbf{Transform}}}
            & \makecell[c]{Discrete Fourier Transform} & \makecell[c]{$X[k] = \sum_{n=0}^{L-1} x[n] e^{-j\frac{2\pi}{L}kn}$} &
            \makecell[l]{
                \textcolor{green}{\checkmark} Ideal for analyzing stationary signals \\
                \textcolor{green}{\checkmark} Efficient in capturing global frequency characteristics \\
                \textcolor{red}{$\times$} Ineffective for non-stationary or time-varying signals \\
            } \\
            \cmidrule(r){2-4}
            & \makecell[c]{Continuous Fourier Transform} &
            \makecell[c]{$X(f) = \int_{-\infty}^{\infty} x(t)e^{-j2\pi ft} dt$} &
            \makecell[l]{
                \textcolor{green}{\checkmark} Suitable for theoretical analysis of continuous signals \\
                \textcolor{green}{\checkmark} Offers a complete and continuous frequency spectrum \\
                \textcolor{red}{$\times$} Impractical for analyzing discrete or sampled data \\
            } \\
            \cmidrule(r){2-4}
            & \makecell[c]{Fast Fourier Transform} & $X[k] = \frac{1}{\sqrt{N}} \sum_{n=0}^{N-1} (x[n] \cdot e^{-j \frac{2\pi}{N}kn})$ &
            \makecell[l]{
                \textcolor{green}{\checkmark} Highly efficient with $\mathcal{O}(N \log N)$ complexity \\
                \textcolor{green}{\checkmark} Widely applied in real-time signal processing \\
                \textcolor{red}{$\times$} Inherits the same limitations as DFT \\
            } \\

        \midrule
        \multirow{5}{*}{\makecell[c]{\textbf{Wavelet} \\ \textbf{Transform}}}
            & \makecell[c]{Discrete Wavelet Transform} & $D(a, b) = \frac{1}{\sqrt{b}} \sum_{m=0}^{p-1} f[t_m] \psi \left( \frac{t_m - a}{b} \right)$ &
            \makecell[l]{
                \textcolor{green}{\checkmark} Captures both time and frequency information \\
                \textcolor{green}{\checkmark} Well-suited for analyzing non-stationary signals \\
                \textcolor{red}{$\times$} Performance depends on the choice of wavelet basis \\
                \textcolor{red}{$\times$} Can be computationally intensive for large-scale data \\
            } \\
            \cmidrule(r){2-4}
            & Continuous Wavelet Transform & \makecell[c]{$F(\tau, s) = \frac{1}{\sqrt{|s|}} \int_{-\infty}^{\infty} f(t) \psi^* \left( \frac{t - \tau}{s} \right) dt$} &
            \makecell[l]{
                \textcolor{green}{\checkmark} Delivers continuous time-frequency representation \\
                \textcolor{green}{\checkmark} Effective for analyzing complex non-stationary signals \\
                \textcolor{red}{$\times$} Computationally demanding and resource-intensive \\
                \textcolor{red}{$\times$} Produces redundant data due to continuous scaling \\
            } \\

        \midrule
        \multirow{4}{*}{\makecell[c]{\textbf{Laplace} \\ \textbf{Transform}}}
            & \makecell[c]{Unilateral Laplace Transform} & \makecell[c]{$F(s) = \mathcal{L}\{f(t)\} = \int_0^\infty f(t)e^{-st} dt$} &
            \makecell[l]{
                \textcolor{green}{\checkmark} Widely used in control theory and differential equations \\
                \textcolor{green}{\checkmark} Facilitates analysis of system stability \\
                \textcolor{red}{$\times$} Less applicable in conventional time series analysis \\
            } \\
            \cmidrule(r){2-4}
            & Bilateral Laplace Transform &
            \makecell[c]{$F(s) = \int_{-\infty}^{\infty} f(t)e^{-st} dt$} &
            \makecell[l]{
                \textcolor{green}{\checkmark} Extends the Fourier Transform\\
                \textcolor{green}{\checkmark} Valuable in engineering and systems analysis \\
                \textcolor{red}{$\times$} Involves complex computations \\
            } \\

        \midrule
        \multirow{4}{*}{\makecell[c]{\textbf{Other} \\ \textbf{Transforms}}}
            & \makecell[c]{Empirical Mode Decomposition} & \makecell[c]{$x(t) = \sum_{j=1}^n c_j(t) + r_n(t)$} &
            \makecell[l]{
                \textcolor{green}{\checkmark} Suitable for non-stationary signals \\
                \textcolor{green}{\checkmark} Decomposes signals into interpretable IMFs \\
                \textcolor{red}{$\times$} Sensitive to noise and end effects \\
            } \\
            \cmidrule(r){2-4}
            & Hilbert-Huang Transform &
            \makecell[c]{$Y(t) = \mathcal{H}[X](t) = \frac{1}{\pi} \int_{- \infty}^{+\infty} \frac{X(s)}{t-s} ds$} &
            \makecell[l]{
                \textcolor{green}{\checkmark} Provides high-resolution analysis \\
                \textcolor{green}{\checkmark} Captures instantaneous frequency and amplitude\\
                \textcolor{red}{$\times$} Relies on quality of EMD \\
            } \\
        \bottomrule[1.2pt]
        \end{tabular}}
    \vspace{-0.1in}
\end{table*}

\subsubsection{Fourier Transform}

The Fourier transform~\cite{nussbaumer1982fast}  serves as a cornerstone mathematical tool for dissecting the frequency composition of signals. It decomposes a time-domain signal into its underlying sinusoidal components, thereby revealing the frequencies present and their corresponding magnitudes. Widely used variants include the Discrete Fourier Transform, Continuous Fourier Transform, and Fast Fourier Transform.

For discrete-time signals, the Discrete Fourier Transform (DFT) provides a discrete frequency representation. The DFT of a sequence $x[n]$ of length $N$ is defined as:
\begin{equation}
\begin{aligned}
    X[k] = \sum_{n=0}^{N-1} x[n] e^{-j\frac{2\pi}{N}kn}, \quad j = \sqrt{-1}
\end{aligned}
\end{equation}
where $X[k]$ denotes the discrete frequency spectrum, and $k$ represents the discrete frequency index. The original time-domain signal can be reconstructed via the Inverse Discrete Fourier Transform (IDFT):
\begin{equation}
\begin{aligned}
x[n] = \frac{1}{N} \sum_{k=0}^{N-1} X[k] e^{j\frac{2\pi}{N}kn}
\end{aligned}
\end{equation}

The Continuous Fourier Transform (CFT) for a continuous-time signal $x(t)$ is defined as follows:
\begin{equation}
    \begin{aligned}
        X(f) = \int_{-\infty}^{\infty} x(t)e^{-j2\pi ft} dt, \quad j = \sqrt{-1}
    \end{aligned}
\end{equation}
where $X(f)$ is the continuous frequency spectrum $x(t)$ is the continuous-time signal. $f$ is the continuous frequency variable. $j$ is the imaginary unit ($\sqrt{-1}$). The inverse CFT reconstructs the time-domain signal from its frequency representation:
\begin{equation} \label{equation5}
    \begin{aligned}
        x(t) = \int_{-\infty}^{\infty} X(f)e^{j2\pi ft} df, \quad j = \sqrt{-1}
    \end{aligned}
\end{equation}

Computationally, a direct DFT calculation exhibits $\mathcal{O}(N^2)$ complexity. The fast Fourier transform (FFT) algorithm offers a remarkable improvement, reducing this complexity to $\mathcal{O}(N \log N)$. This substantial efficiency gain makes the FFT an indispensable tool in various domains, including signal processing, telecommunications, and image analysis, where swift frequency-domain analysis is paramount. The FFT achieves its efficiency through a recursive decomposition of the DFT computation, exploiting inherent symmetries in the trigonometric functions involved. In practical applications, to simplify calculations and reduce redundancy, the FFT is typically employed to implement the DFT~\cite{nussbaumer1982fast}. For complex signals, let $x[n]$ represent the original discrete signal with a length of $N$. We use orthogonal normalization to convert the signal to the frequency-domain:\begin{equation} \label{equation6}
    \begin{aligned}
        X[k] = \frac{1}{\sqrt{N}} \sum_{n=0}^{N-1} (x[n] \cdot e^{-j \frac{2\pi}{N}kn}), \quad k=0,1,\cdots,N-1
    \end{aligned}
\end{equation}

The signal can also be converted back to the time domain using an Inverse Fast Fourier Transform (IFFT):\begin{equation}
    \begin{aligned}
        x[n] = \frac{1}{\sqrt{N}} \sum_{k=0}^{N-1} (X[k] \cdot e^{j \frac{2\pi}{N}kn}), \quad n=0,1,\cdots,N-1
    \end{aligned}
\end{equation}

Orthogonal normalization ensures the preservation of signal energy, making it independent of signal length:\begin{equation}
    \begin{aligned}
        x[n] = \text{IFFT}(\text{FFT}(x[n]))
    \end{aligned}
\end{equation}

Despite unparalleled efficacy in stationary frequency decomposition, Fourier methods exhibit an inherent resolution constraint: the uncertainty principle fundamentally limits simultaneous temporal and spectral precision. This manifests as compromised event localization when analyzing transient phenomena where timing fidelity proves critical. Consequently, wavelet transforms emerged to overcome this spatiotemporal trade-off through adaptive basis functions, which will be discussed in the next section.

\subsubsection{Wavelet Transform}
\label{wavelet transformer}
The wavelet transform~\cite{meyer1989wavelets} is a mathematical tool used in signal processing that allows for the decomposition of a signal in the time-frequency domain. Unlike the Fourier transform, the wavelet transform can reveal both the frequency characteristics of a signal and the time distribution of these frequency components, making it particularly effective for analyzing non-stationary signals. According to the computation methods, wavelet transforms can be divided into the continuous wavelet transform and discrete wavelet transform.

The continuous wavelet transform (CWT) computes wavelet coefficients by analyzing a signal across different frequencies and time positions, providing a detailed energy distribution at the expense of high computational cost. The following equation shows the CWT~\cite{leung2021financial}, which decomposes a continuous signal $f(t)$ using a mother wavelet function $\psi$ and its complex conjugate $\psi^*$ into frequency components localized in time: $F(\tau, s) = \frac{1}{\sqrt{\vert s \vert}} \int_{-\infty}^{\infty} f(t) \psi^* \left( \frac{t - \tau}{s} \right) dt$,
where $\tau$ is the translation parameter, determining the position of the wavelet along the time axis, and $s$ is the scale parameter, which controls the width of the wavelet and thus affects the frequency resolution. This continuous transform enables the examination of the signal's frequency components in a time-localized manner, making it suitable for analyzing non-stationary signals.

The discrete wavelet transform (DWT) performs multi-scale decomposition of a signal, separating it into different frequency bands while maintaining both frequency and time localization. Unlike the CWT, the DWT is computationally efficient and suitable for digital signal processing. The following equation illustrates the DWT, which decomposes a discrete signal $f[t_m]$ into wavelet coefficients $D(a, b)$ based on discrete scaling and translation steps $D(a, b) = \frac{1}{\sqrt{b}} \sum_{m=0}^{p-1} f[t_m] \psi \left( \frac{t_m - a}{b} \right)$,
where $a$ is the translation parameter, controlling the position of the wavelet on the time axis, and $b$ is the scale parameter, determining the width of the wavelet, which affects the frequency resolution. This transformation enables the decomposition of the signal $f[t_m]$ into components at various scales and positions.

\subsubsection{Laplace Transform}

The Laplace transform~\cite{schiff2013laplace} is a powerful integral transform widely used in the analysis of linear time-invariant (LTI) systems. It converts a function of time, often representing a system's input or output signal, into a function of a complex frequency variable, $s$. This transformation simplifies the analysis of differential equations, frequently encountered in describing LTI system behavior. The key benefit is the conversion of differential equations into algebraic equations, which are significantly easier to solve. The two primary types of Laplace transforms are unilateral and bilateral. The unilateral Laplace transform is particularly useful in control systems and the analysis of differential equations, as it aids in system stability analysis. However, it is less commonly used in traditional time-series analysis. The bilateral Laplace transform generalizes the Fourier transform and is used primarily in engineering and systems analysis, but it is computationally intensive due to its broader scope.

Specifically, the unilateral Laplace transform of a function $f(t)$, defined for $t \ge 0$, is given by:
\begin{equation}\label{eq:laplace}
    F(s) = \mathcal{L}\{f(t)\} = \int_0^\infty f(t)e^{-st} dt
\end{equation} where $F(s)$ represents the Laplace transform of $f(t)$. $s$ is a complex frequency variable, often expressed as $s = \sigma + j\omega$, with $\sigma$ representing the real part (related to exponential decay) and $\omega$ the imaginary part (related to frequency). The integral's lower limit of 0 reflects the typical application to causal signals (signals that are zero for $t < 0$).

The inverse Laplace transform recovers the original time-domain function $f(t)$ from its Laplace transform $F(s)$:
\begin{equation}
f(t) = \mathcal{L}^{-1}\{F(s)\} = \frac{1}{2\pi j} \int_{\sigma - j\infty}^{\sigma + j\infty} F(s)e^{st} ds
\end{equation}
This integral is a complex line integral along a vertical line in the complex $s$-plane, with $\sigma$ chosen to ensure convergence.  The inverse transform is frequently computed using tables of Laplace transforms and techniques like partial fraction decomposition.

\subsubsection{Other Transforms} There are other frequency-domain approaches, such as Empirical Mode Decomposition (EMD)~\cite{zeiler2010empirical} and the Hilbert-Huang Transform (HHT)~\cite{huang2014hilbert}. EMD is suitable for nonlinear signal analysis, which adaptively represents the local characteristic of the given signal. Based on EMD, any complicated signal can be decomposed into a finite and often small number of Intrinsic Mode Functions (IMFs), which have simpler frequency components and stronger correlations~\cite{cheng2014novel}, thus are easier and more accurate to forecast. Specifically, for any given time series $x(t)$ observed over a period of time $[0, T]$, we decompose it in an iterative way into a finite sequence of oscillating components $c_1(t), \cdots, c_n(t)$, plus a non-oscillatory trend called the residue term $r_n(t)$. Precisely, we have $x(t) = \sum_{j=1}^n c_j(t) + r_n(t)$.
In addition, the HHT method can decompose any time series into oscillating components with nonstationary amplitudes and frequencies using the EMD~\cite{leung2021financial}. This fully adaptive method provides a multiscale decomposition for the original time series, which gives richer information about the time series. The instantaneous frequency and instantaneous amplitude of each component are later extracted using the Hilbert transform. For any function in time $X(t)$, the Hilbert transform is given by\begin{equation}
    \begin{aligned}
        Y(t) = \mathcal{H}[X](t) := \frac{1}{\pi} \int_{- \infty}^{+\infty} \frac{X(s)}{t-s} ds
    \end{aligned}
\end{equation} where the improper integral is defined as the Cauchy principal value. The decomposition onto different timescales also allows for reconstruction up to different resolutions, providing a smoothing and filtering tool that is ideal for noisy financial time series.

\subsubsection{Comparative Operational Domains} Time-frequency analysis techniques have domain-specific advantages. Fourier transforms are optimal for stationary signals, offering $O(N \log N)$ computational efficiency, but are limited by fixed time-frequency resolution and spectral leakage, making them suitable for steady-state applications like communication system analysis. Wavelet transforms, on the other hand, excel at analyzing non-stationary signals with adaptive time-frequency localization, making them ideal for transient detection in applications such as EEG analysis or mechanical fault diagnosis, though their performance is sensitive to the choice of basis functions. Laplace transforms provide unique advantages in system dynamics modeling by converting differential equations, reducing error accumulation in long sequences, and detecting operational shifts, but are constrained to linear time-invariant systems like control circuits. Adaptive decompositions handle nonlinearity through data-driven methods, avoiding basis mismatch and cross-term interference, making them especially valuable for complex signals such as financial time series or turbulence data.

\subsection{Studies on Classical Methods}  The previous section discussed various signal transform techniques, each with distinct trade-offs in time-frequency resolution and adaptability to non-stationary signals. Despite their differences in complexity and application, these methods have been crucial in traditional signal analysis. This section reviews classical studies that effectively utilized these techniques.

\subsubsection{Studies based on Fourier Transform} DFT captures global frequency components efficiently but loses time-domain information and cannot handle non-stationary signals. CFT provides a continuous spectrum but is impractical for discrete signals. FFT offers fast computation with $\mathcal{O}(N \log N)$ complexity, making it suitable for real-time applications, though it shares DFT's limitations. Each method has unique advantages and trade-offs depending on the signal characteristics and application requirements. For example, NFT~\cite{koren2024interpretable} demonstrates how integrating multi-dimensional Fourier transforms with deep learning frameworks can enhance both predictive accuracy and interpretability in financial forecasts. Similarly, approaches like FourNet~\cite{du2023fourier} employ Fourier-based neural networks to approximate transition densities in complex financial models, providing rigorous error bounds and robust performance on diverse stochastic processes. Pastnet~\cite{wu2024pastnet} addresses these challenges by employing spectral methods that integrate trainable neural networks with Fourier-based a priori spectral filters, transforming raw data into frequency-domain representations where the Fourier coefficients capture the intrinsic periodic features of the system, thereby enabling the model to achieve state-of-the-art performance in both weather forecasting and traffic prediction. Besides, LPR~\cite{chen2024laplacian} synergistically combines the circulant matrix nuclear norm with Laplacian kernelized temporal regularization to yield a unified framework via FFT in log-linear time complexity, accurately imputing diverse traffic time series behaviors and reconstructing sparse vehicular speed fields.

Other methods, such as the Short-Time Fourier Transform (STFT)~\cite{yao2019stfnets} and the Fractional Fourier Transform (FrFT)~\cite{kocc2022fractional}, address specific needs. STFT enables frequency analysis with time localization, commonly used in speech and signal processing, but involves a trade-off between time and frequency resolution. FrFT generalizes the Fourier transform for non-stationary signals but is more complex and computationally intensive. To support graph data, extensions like the spectral graph Fourier transform~\cite{defferrard2016convolutional} and wavelet graph transform~\cite{xu2019graph} provide localized frequency analysis for irregularly structured time series. However, they require graph construction and eigen decomposition, and are more complex than traditional methods.

\subsubsection{Studies based on Wavelet Transform} Recent works have leveraged wavelet transforms for diverse tasks, such as optimizing time-frequency representations through non-linear filter-bank transformations~\cite{cosentino2020learnable}, isolating periodic components using the maximal overlap DWT~\cite{wen2021robustperiod}, and integrating wavelet methods into deep learning frameworks to capture both frequency and time-domain features~\cite{yang2023waveform}. Moreover,~\citep{liang2024waverora} introduced wavelet-based frameworks that leverage time-frequency features to enhance forecasting efficiency and accuracy.

Furthermore, some methods combine the strengths of Fourier and wavelet transforms to improve time-series analysis. Zhou \textit{et al.}~\cite{zhou2022fedformer} proposed FEDformer, which integrates Fourier transform for capturing global patterns and wavelet transform for modeling local structures, achieving a balance between accuracy and computational efficiency. Liu \textit{et al.}~\cite{liu2024wftnet} introduced WFTNet, which incorporates both transforms with a periodicity-weighted coefficient to adaptively balance their contributions. These hybrid approaches effectively exploit the complementary strengths of Fourier and wavelet transforms, setting new benchmarks in long-term time-series forecasting.

\begin{figure*}[htb!]
    \vspace{-0.1in}
    \centering
    \includegraphics[width=0.90\linewidth]{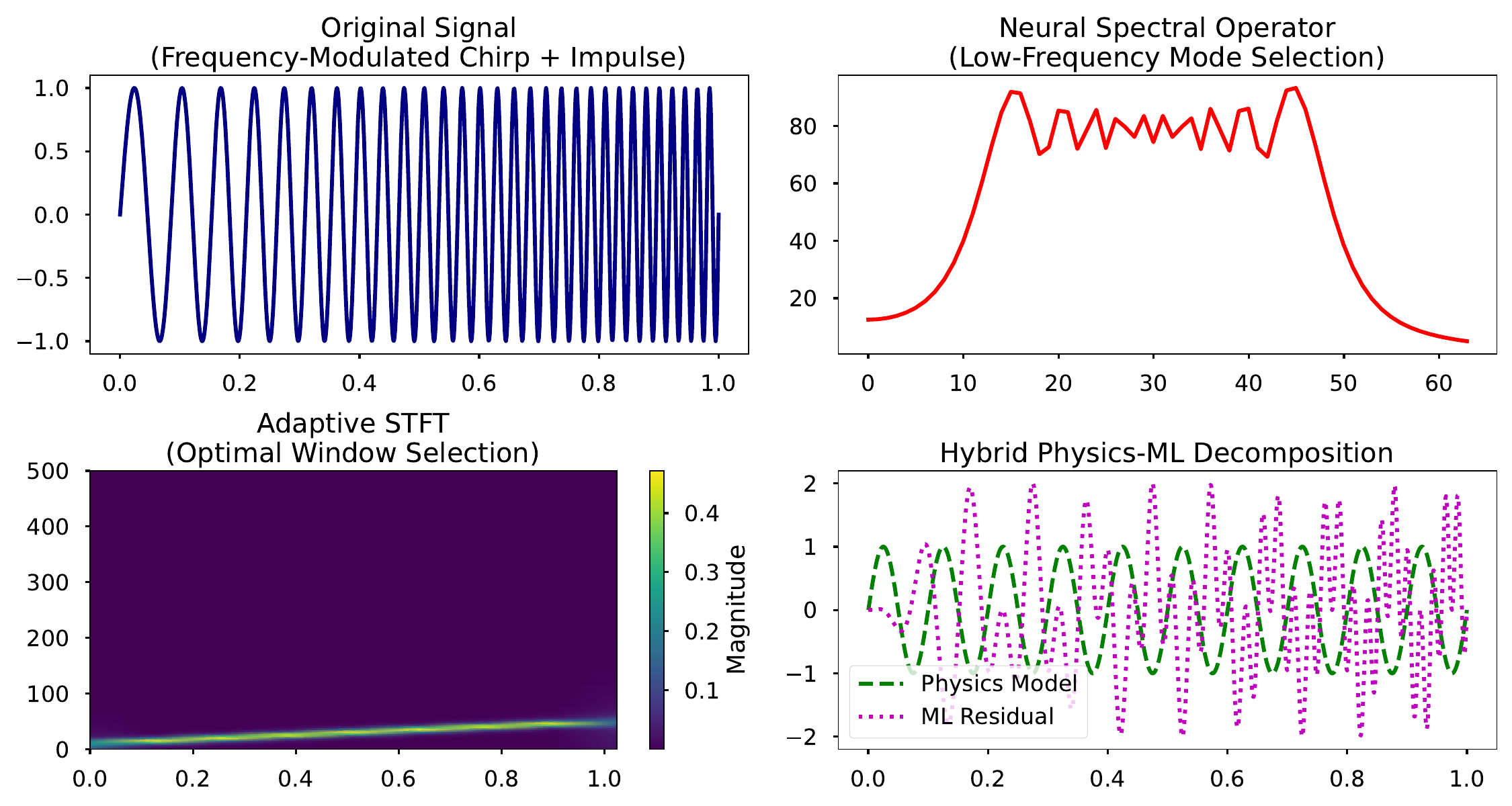}
    \caption{Time series analysis in different kinds of modern frequency transform methods.}
    \label{fig:modern_methods}
    \vspace{-0.2in}
\end{figure*}

\subsubsection{Studies based on Laplace Transform} While the Laplace transform has a wide range of applications in machine learning, its direct application to time-series data remains limited, likely due to challenges in integrating it with complex temporal structures. For instance,~\citep{ambhika2024time} proposed a hybrid model combining Laplace transform-based deep recurrent neural networks with long short-term memory networks for time-series prediction. Similarly,~\citep{chen2024laplacian} and~\citep{shu2024low} leveraged Laplacian transforms for traffic time-series imputation, using methods like low-rank completion, Laplacian kernel regularization, and FFT. These studies highlight the Laplace transform’s potential in improving time-series modeling by addressing challenges in data representation and computational efficiency.

\subsubsection{Studies based on Other Transforms} There are many time series analysis studies based on Empirical Mode Decomposition (EMD) and the Hilbert-Huang Transform (HHT). EMD is a data-driven method that decomposes a non-linear and non-stationary time series into a finite set of Intrinsic Mode Functions (IMFs), which represent simple oscillatory modes embedded in the time series data~\cite{cheng2014novel, yang2019hybrid,buyukcsahin2019improving}. The Hilbert-Huang Transform extends this by applying the Hilbert Transform to the IMFs to obtain instantaneous frequency and amplitude, offering a highly adaptive time-frequency representation~\cite{hao2015analysis,leung2021financial,dezhkam2023forecasting}. Unlike traditional transforms such as Fourier and Laplace, which assume linearity and stationarity, or Wavelet Transform, which uses predefined basis functions, EMD and HHT are empirical and adaptive, making them particularly suitable for analyzing real-world, complex time series signals with dynamic spectral characteristics.

\subsubsection{Summary} The classical frequency transform landscape reveals a complementary ecosystem where each method addresses specific signal characteristics and application requirements: Fourier transforms serve as the foundation for stationary signal processing with their computational efficiency and global frequency analysis capabilities, wavelet transforms overcome temporal localization limitations through adaptive time-frequency resolution to become essential for non-stationary signal analysis, Laplace transforms provide specialized capabilities in system dynamics and control theory, while empirical methods like EMD and HHT introduce data-driven adaptability for complex nonlinear signals. Contemporary research trends show that hybrid methodologies like FEDformer and WFTNet successfully combine multiple transform techniques to achieve superior performance, while integration with deep learning frameworks (as demonstrated by NFT, FourNet, and Pastnet) further illustrates how classical mathematical foundations can be enhanced through modern computational approaches. This convergence not only addresses the fundamental trade-offs inherent in individual transform methods, resolution-efficiency balance and global versus local analysis requirements, but also points toward unified frameworks that can automatically select optimal transform strategies based on signal characteristics, ultimately preserving theoretical rigor while achieving practical flexibility demanded by modern applications.

\section{Recent Methodological Advances}\label{sec:current}

Recent advances in spectral operator learning combine neural networks with mathematical foundations. Figure~\ref{fig:modern_methods} highlights two paradigm-shifting approaches: (1) Fourier Neural Operators enabling $\mathcal{O}(N \log N)$ PDE solutions via spectral filters, and (2) Wavelet Neural Operators capturing transient phenomena through adaptive time-frequency localization. Hybrid methods integrate physical constraints via composite loss functions (Physics-Informed FNO) or stochastic terms (Neural SPDEs). These techniques outperform traditional methods in accuracy and efficiency across climate modeling, turbulent flows, and financial forecasting. For clear illustration, we provide a comparison of them in Figure~\ref{fig:modern_methods}.

This comparison showcases cutting-edge approaches for analyzing non-stationary signals. The Neural Spectral Operator (top right) demonstrates how machine learning can intelligently preserve dominant frequency modes while filtering noise - a technique particularly effective for signals with underlying physical laws. The Adaptive STFT (bottom left) automatically adjusts its time-frequency resolution, simultaneously capturing both the chirp's evolving frequency (visible as curved energy bands) and the transient impulse at t=0.5s (vertical spike). Most innovatively, the Hybrid Physics-ML decomposition (bottom right) separates the signal into a physics-based component (representing known system dynamics) and a data-driven residual, offering interpretable modeling for complex systems. Together, these methods overcome classic limitations: where traditional Fourier analysis fails to localize transient events and standard wavelets require manual scale selection, these adaptive techniques automatically optimize their representations while maintaining physical plausibility. Such approaches are revolutionizing applications from predictive maintenance (detecting equipment faults) to climate science (analyzing extreme weather patterns), where both precise timing and frequency content are critical.

\subsection{Spectral Neural Operators}
\textbf{Fourier Neural Operators (FNOs)} represent a groundbreaking architecture for learning mappings between function spaces, particularly effective for solving partial differential equations (PDEs) and modeling complex physical systems. The core innovation lies in operating directly in the frequency domain to capture global patterns efficiently. The mechanism of FNOs is shown as follows:
\begin{equation}
(K_\theta v)(x) = \mathcal{F}^{-1}\left(R_\theta(k) \cdot \mathcal{F}(v)(k)\right)(x)
\end{equation}
where $R_\theta(k) \in \mathbb{C}^{d \times d}$ is the learned frequency-domain kernel. $\mathcal{F}$ denotes the Fourier transform. The computational complexity of FNOs is $\mathcal{O}(N \log N)$ via FFT. The key architectures are formed as follows: (1) Input Transformation: The process begins by lifting the input function $v(x)$ to a higher-dimensional representation space through a shallow neural network P via $v'(x) = P(v(x))$. This initial transformation enhances the model's capacity to represent complex features. (2) Spectral Processing
The system then applies a Fourier transform $\mathcal{F}$ to decompose the input into its frequency components via
$\tilde{v}(k) = \mathcal{F}({v'(x)})$. In this spectral domain, the model applies learnable complex-valued weights $R_\theta(k)$ to modulate different frequency bands via $u(k) = R_\theta(k)\tilde{v}(k)$.

\textbf{Wavelet Neural Operators (WNOs)} are a powerful extension of neural operators that leverage wavelet transforms to efficiently capture multi-scale, localized patterns in data, making them particularly effective for modeling non-stationary and transient phenomena in time series, PDEs, and high-dimensional dynamical systems. The main idea of WNOs is shown as follows:
\begin{equation}
W_\psi f(a,b) = \frac{1}{\sqrt{a}}\int_{-\infty}^\infty f(t)\psi^*\left(\frac{t-b}{a}\right)dt
\end{equation}
where $\psi$ is a chosen admissible mother wavelet, while scale- and translation-dependent coefficients are learned by the operator. $a$ is the scale parameter, which controls frequency. $b$ is the translation parameter which controls localization. The process of inverse wavelet reconstruction is shown as follows:
\begin{equation}
v_{\text{out}}(x) = \frac{1}{C_\psi} \int_{0}^{\infty} \int_{-\infty}^{\infty} W_\psi v(a,b) \, \psi_{a,b}(x) \, \frac{da \, db}{a^2}
\end{equation}
where $C_\psi$ is a normalization constant. The computational time complexity is $\mathcal{O}(N)$.

\subsection{Adaptive Time-Frequency Analysis}

The \textbf{learned wavelet transform} represents an adaptive approach to time-frequency analysis where the wavelet parameters are optimized during training. The parametric mother wavelet is defined as $\psi_\theta(t) = \frac{1}{\sqrt{a_\theta}} \psi\left(\frac{t-b_\theta}{a_\theta}\right)$
where $a_\theta$ controls the scale parameter, determining the frequency bandwidth.
$b_\theta$ governs the position parameter, localizing the wavelet in time.
$\theta$ denotes the learnable parameters of the system. In this formulation, $a_\theta$ and $b_\theta$ are trainable scale and translation parameters applied to a fixed admissible mother wavelet $\psi$ (e.g., Morlet or Mexican hat). We enforce $a_\theta>0$ (for example, by a softplus parameterization), so dilation is well defined. Learning the samples or functional form of $\psi$ itself is a different design choice and requires explicit admissibility and zero-mean constraints. The parameters are optimized through backpropagation using the chain rule $\frac{\partial \mathcal{L}}{\partial \theta} = \frac{\partial \mathcal{L}}{\partial W_\psi} \cdot \frac{\partial W_\psi}{\partial \theta}$
This enables multi-resolution learning via different frequency bands and signal-adaptive localization to focus on important temporal regions.
The \textbf{Neural Hilbert Spectrum} extends traditional Hilbert-Huang analysis by incorporating deep learning:
\begin{equation}
H(\omega,t) = \text{ReLU}(\text{CNN}({IMF_k(t)}_{k=1}^K)) \cdot \delta(\omega-\omega_k(t))
\end{equation}
where CNN processing refines the intrinsic mode functions (IMFs).
ReLU activation ensures non-negative energy values.
The Dirac delta function $\delta(\omega-\omega_k(t))$ assigns energy to instantaneous frequencies.

\subsection{Hybrid Physics-ML}
The \textbf{Physics-Informed FNO} combines the strengths of neural operators with physical constraints through a composite loss function:
\begin{equation}
    \mathcal{L} = \underbrace{|u_\theta - u^*|^2}_{\text{data fidelity}} + \lambda \underbrace{|\partial_t u_\theta - \mathcal{N}[u_\theta]|^2}_{\text{physics constraint}}
\end{equation}
where $|u_\theta - u^*|^2$ denotes the data-fidelity term: the discrepancy between model predictions $u_\theta$ and reference observations or high-fidelity simulations $u^*$. $|\partial_t u_\theta - \mathcal{N}[u_\theta]|^2$ denotes the physics residual, and $\mathcal{N}[\cdot]$ represents the governing differential operator (e.g., Navier--Stokes or a wave operator).
$\lambda$ controls the trade-off between data fitting and physics adherence, with larger $\lambda$ enforcing stronger satisfaction of the governing PDE.

\textbf{Neural Stochastic PDEs} extend neural operators to stochastic dynamical systems:
\begin{equation}
\begin{aligned}
du_t &= \Big[\underbrace{\mathcal{F}_\theta(u_t)}_{\text{neural drift}} + \underbrace{\nabla \cdot (D_\theta \nabla u_t)}_{\text{diffusion}} \Big]dt + \underbrace{\sigma_\theta(u_t)dW_t}_{\text{stochastic forcing}}
\end{aligned}
\end{equation}
where $\mathcal{F}_\theta(u_t)$ denotes the drift term and learns deterministic dynamics via neural network (often FNO or CNN).
$\nabla \cdot (D_\theta \nabla u_t)$ denotes the diffusion term.
$D_\theta$ learns a diffusivity tensor; positive definiteness supports physically meaningful diffusion, although stability also depends on the remaining dynamics and numerical discretization.
$\sigma_\theta(u_t)dW_t$ is the stochastic term. $\sigma_\theta$ is adopted to learn volatility function (state-dependent noise).
$dW_t$ denotes the Wiener process increments modeling unresolved scales.

\subsection{Studies on Recent Frequency Methods}
Recent advancements in frequency-domain analysis have revolutionized time series processing by enabling more robust feature extraction, noise resilience, and interpretability. Modern techniques leverage spectral transformations - such as wavelet decompositions, short-time Fourier transforms, and Hilbert-Huang transforms - to capture multi-scale patterns, transient events, and non-stationary behaviors in complex datasets. Innovations like adaptive wavelet learning, neural spectrum analyzers, and physics-informed frequency embeddings further enhance discriminative power for tasks like anomaly detection, classification, and forecasting. These methods excel in scenarios where traditional time-domain models struggle, such as analyzing biomedical signals, financial volatility, or industrial sensor data with inherent periodicity or abrupt changes. Recent studies also integrate frequency features with deep learning architectures (e.g., attention mechanisms) to improve cross-domain generalization. By transforming temporal data into interpretable spectral representations, these approaches bridge the gap between signal processing theory and cutting-edge machine learning, offering new insights for both theoretical and applied research.
\subsubsection{Studies based on Spectral Neural Operators (SNOs)}
SNOs have emerged as a powerful framework for learning solution operators of complex differential equations and spatiotemporal systems. By operating directly in the frequency domain, these architectures efficiently capture global patterns while maintaining resolution independence. Recent work spans distinct settings, including PDE/operator learning~\cite{cao2025spectral,oommen2025integrating,tiwari2025cono,qin2024toward,michalowska2024neural,du2023neural,bahmani2025neural}, time-series forecasting and chaotic dynamics~\cite{damashifting,long2024time,wang2024chaotic}, and optical-fiber propagation modeling~\cite{zhang2025fourier}. These studies use spectral representations for different objectives and protocols and should not be interpreted as results on a common time-series benchmark. Several of these methods use Fourier or wavelet transforms with reported
$\mathcal{O}(N \log N)$ complexity, enabling scalable modeling of partial differential equations (PDEs), turbulent flows, and high-dimensional dynamical systems. To be specific,
~\citet{cao2025spectral} propose an innovative learning framework, namely spectral-refiner, that significantly enhances Fourier Neural Operators (FNOs) through two fundamental advancements. First, spectral-refiner develops a novel spatiotemporal extension that generalizes FNO architectures to operate effectively in Bochner spaces, enabling unprecedented arbitrary-length temporal super-resolution capabilities. Second, spectral-refiner introduces a paradigm-shifting training methodology that synergizes deep learning with numerical PDE techniques, replacing conventional end-to-end approaches with a more efficient two-stage process. The framework first conducts brief preliminary training on Navier-Stokes equation simulations, then implements specialized fine-tuning of our enhanced spectral convolution layer. This refinement phase incorporates several breakthroughs: complete retention of frequency components (eliminating traditional truncation), a pioneering convex optimization scheme using negative Sobolev norms, and precise error quantification through Parseval-identity-based estimators. Another recent study~\cite{bahmani2025neural} proposes a novel approach that adopts the spectral expansion framework of polynomial chaos expansion (PCE) while replacing traditional polynomial basis functions with neural network (NN)-based alternatives to enhance expressivity. This PCE method constructs these NN-parameterized basis functions sequentially, enforcing orthogonality with respect to the empirical data distribution without making prior assumptions about the joint distribution of random variables - whether dependent or independent, or their marginal distributions. Unlike conventional PCE methods, this PCE directly operates on joint stochastic variables, eliminating the need for tensor product structures or independence assumptions between variables. This provides greater flexibility for modeling complex stochastic systems while simultaneously simplifying implementation compared to standard PCE techniques that rely on tensor product formulations for handling random vectors.

\subsubsection{Studies based on Adaptive Time-Frequency Analysis}
Time series data, characterized by their temporal evolution, often exhibit complex non-stationary behaviors where frequency components vary over time. Traditional Fourier analysis, while powerful, assumes stationarity and fails to capture such dynamic spectral features. To address this limitation, adaptive time-frequency analysis (TFA) has emerged as a critical tool for revealing the time-localized frequency content of signals. Unlike fixed-resolution methods like the short-time Fourier transform (STFT), adaptive TFA techniques automatically adjust their time-frequency resolution to match signal characteristics. Approaches such as wavelet transforms, empirical mode decomposition (EMD), and synchrosqueezing transforms provide high-resolution representations of transient phenomena, oscillatory patterns, and non-linear trends. The cited studies address different applications, including medical-signal denoising and sleep staging~\cite{chao2025self,liang2025adaptive}, financial forecasting~\cite{zhang2025fan}, sequential recommendation~\cite{lu2025filtering}, and general adaptive time-frequency analysis~\cite{li2025faith,xu2025local,luo2024time,ye2024atfnet,huang2024adaptive,leiber2024adaptive,warion2024class}; their results are application-specific rather than directly comparable. ~\citet{chao2025self} propose a self-adjusting frequency decomposition framework with a learnable regularization weighting strategy. Their medical-signal experiments report improved noise robustness while preserving signal morphology. A non-peer-reviewed preprint~\cite{zhang2025fan} proposes a Frequency-Adaptive Normalized (FAN) framework for stock-market time-series forecasting. Its architecture uses parallel adaptive frequency normalization and residual-learning paths to model periodic structure and local variation.

\subsubsection{Studies based on Hybrid Physics-Machine Learning (ML) Methods}

A conventional two-stage model first estimates system-derived features or states and then trains a predictor. Hybrid physics-ML methods may instead place governing structure inside the architecture, objective, or learned dynamics and can be trained end to end. They complement spatio-temporal causal modeling rather than subsume it. Lindeberg's time-causal and time-recursive scale-space constructions~\cite{lindeberg2023time,lindeberg2025time} are theoretical signal representations, whereas Azad et al.~\cite{azad2024vision} present a research vision for spatio-causal situation awareness; these roles are not equivalent forms of experimental evidence.

Hybrid physics-ML methods have emerged as a powerful paradigm for time-series analysis, combining the interpretability of physical models with the flexibility of data-driven approaches. These methods complement, rather than subsume, two-stage pipelines and spatio-temporal causal modeling; related time-causal and time-recursive scale-space representations~\cite{lindeberg2023time,lindeberg2025time} and spatio-causal situation awareness~\cite{azad2024vision} enforce physical and causal structure beyond standard PINN-style composite losses. These methods integrate domain-specific differential equations or physical constraints with ML architectures (e.g., neural operators, PINNs) to model complex temporal dynamics while respecting underlying physical laws. In applications ranging from climate modeling to industrial prognostics, hybrid frameworks demonstrate superior performance over purely data-driven models, particularly in scarce-data regimes where physical priors prevent overfitting. Key innovations include physics-guided loss functions that penalize non-physical predictions, coupled architectures where ML corrects residual errors of simplified physics models, and neural differential equations that learn interpretable system parameters. Such methods~\cite{jiao2024hybrid,qi2025coupled,wang2024hybrid,christopoulos2025towards,liu2022hybrid,abbasi2024physics,hsieh2024hybrid,lin2023systematic,lin2024stress,lin2023sampling} not only achieve higher accuracy in forecasting non-stationary, noisy signals but also provide actionable insights into system behavior, bridging the gap between theoretical understanding and empirical patterns in real-world time-series data. To be specific, ~\citet{abbasi2024physics} propose to integrate machine learning components through domain-specific computational graphs that encode physical constraints. These customized connections: (i) prevent violations of governing physical laws, (ii) enhance feature propagation efficiency, and (iii) maintain interpretability. Extensive testing on various coupled nonlinear systems confirms the approach's versatility in handling complex, multidimensional dynamical processes. Current machine learning implementations frequently demonstrate unstable and unreliable behavior when integrated within coupled climate modeling frameworks, particularly during dynamic interactions with large-scale climate simulations. These limitations become especially pronounced under out-of-distribution climatic conditions. While architectural innovations like climate-invariant moisture transformations, expanded input vectors, and temporal context inclusion have demonstrated some coupled performance improvements, they remain inadequate for robust generalization. The development of advanced feature engineering techniques~\cite{lin2024stress} capable of preventing unphysical extrapolations could significantly enhance the utility of observational data in hybrid physics-ML climate modeling systems.

\section{Benchmarks and Evaluation}
\label{sec:benchmark}

\subsection{Benchmark Datasets}

Table~\ref{tab:benchmark_dataset} provides the statistics and feature details of commonly used datasets, which also include multivariate data, for time series analysis. These datasets cover applications ranging from energy consumption and meteorological indicators to healthcare and anomaly detection, offering a comprehensive foundation for research and model benchmarking.
\begin{itemize}
    \item Electricity Transformer Temperature (ETT)~\citep{zhou2021informer} contains four sub-datasets: ETTm1\&m2 and ETTh1\&h2, collected from two electricity transformers at two stations at different resolutions (15min or 1h). ETT dataset contains multiple series of loads and one series of oil temperatures;
    \item Weather dataset~\cite{wetterstation} contains 21 meteorological indicators for a range of 1 year in Germany;
    \item Electricity (ECL) dataset~\cite{uci_electricity} contains the electricity consumption of clients with each column corresponding to one client;
    \item Traffic dataset~\cite{pems_traffic} contains the occupation rate of freeway system across the State of California;
    \item Exchange dataset~\citep{lai2018modeling} contains the current exchange of eight countries;
    \item Illness (ILI) dataset~\cite{cdc_fluview} contains the influenza-like illness patients in the United States.
    \item The UEA multivariate time-series classification archive~\cite{uea_ucr_tsc} contains datasets spanning diverse application domains and varying numbers of channels and classes.
    \item Server Machine Dataset (SMD)~\cite{su2019omnianomaly} is a 5-week dataset from 28 machines in three groups, designed for anomaly detection with labeled points and contributing dimensions, trained/tested separately.
    \item CausalRivers~\cite{gideon2025causalrivers} provides spatially grounded real-world river time series for large-scale causal discovery benchmarking.
\end{itemize}

\begin{table*}[!ht]
    \vspace{-0.1in}
    \renewcommand\arraystretch{1.5}
    \setlength{\tabcolsep}{8pt}
    \centering
    \caption{A list of commonly used and publicly accessible datasets.}
    \vspace{-0.15in}
    \label{tab:benchmark_dataset}
        \setlength{\tabcolsep}{5pt}
    \resizebox{\textwidth}{!}{
                \begin{threeparttable}
        \begin{tabular}{c|c|c|c|c|c|c|c|c}
        \toprule[1.2pt]
        \textbf{Datasets} & \textbf{Variates}  & \textbf{Pred. Len.} & \textbf{Dataset Size} & \textbf{Freq.} & \textbf{Source} & \textbf{Spatial metadata} & \textbf{Known causal graph} & \textbf{Task} \\
        \midrule[1.2pt]
        ETTm1\&m2 & 7 & $\{96,\ldots,720\}$ & (34465,11521,11521) & 15 mins & Observational & No & No & \makecell[c]{Forecasting} \\
        \midrule
        ETTh1\&h2 & 7 & $\{96,\ldots,720\}$ & (8545,2881,2881) & 1 h & Observational & No & No & \makecell[c]{Forecasting} \\
        \midrule
        Weather & 21 & $\{96,\ldots,720\}$ & (36792,5271,10540) & 10 mins & Observational & No & No & \makecell[c]{Forecasting} \\
        \midrule
        Electricity & 321 & $\{96,\ldots,720\}$ & (370, 140256) & 15 mins & Observational & No & No & \makecell[c]{Forecasting} \\
        \midrule
        Traffic & 862 & $\{96,\ldots,720\}$ & (12185,1757,3509) & 1 h & Observational & Not distributed & No & Forecasting \\
        \midrule
        Exchange & 8 & $\{96,\ldots,720\}$ & (5120,665,1422) & 1 day & Observational & No & No & Forecasting \\
        \midrule
        Illness & 8 & $\{1, 2, 3, 4\}$ & $\approx$ (2600, 500, 1000) & 7 day & Observational & No & No & Forecasting \\
        \midrule
        UEA  & 1 \textasciitilde 1344 & - & Varies & - & Mixed collection & Dataset-dependent & No & Classification \\
        \midrule
        SMD & 38 & - & (566724,141681,708420) & 1 min & Observational & No & No & \makecell[c]{Anomaly Det.} \\
        \midrule
        CausalRivers & 666/494 stations & - & 2019--2023 & 15 mins & Observational & Yes & Yes & Causal discovery \\
        \bottomrule[1.2pt]
        \end{tabular}
        \end{threeparttable}
    }
    \vspace{-0.1in}
\end{table*}
In Table~\ref{tab:benchmark_dataset}, ``spatial metadata'' means that locations or an explicit spatial structure are distributed with the benchmark; ``known causal graph'' means that the benchmark supplies a reference graph for evaluation. We replace the former dataset-level ``linearity'' label because linearity depends on the variables, transformations, scale, and diagnostic used and cannot be assigned reliably as a single intrinsic property. CausalRivers contains observational river-discharge series from 666 eastern-German and 494 Bavarian stations, sampled every 15 minutes from 2019--2023, together with constructed reference graphs~\cite{gideon2025causalrivers}.

\subsection{Comparative Analysis of Approaches}

Tables~\ref{tab:bench_comparision} and~\ref{tab:bench_multidataset} compare values reported in the original papers; we did not re-run the methods under a unified setup. Even when papers use nominally common datasets and horizons, differences in data splits, lookback windows, preprocessing, framework versions, hardware, tuning, and random seeds can affect results~\cite{shahriari2022reproducibility,gundersen2022reproducibility}. The tables should therefore be read as a descriptive synthesis of published evidence, not as a controlled ranking. Future benchmark studies should use a standardized evaluation pipeline and report code versions, splits, preprocessing, seeds, and uncertainty.

On Weather, the published 96-step MSE values include 0.160 for WaveForM~\cite{yang2023waveform}, 0.159 for WaveRoRA~\cite{liang2024waverora}, and 0.143 for FRNet~\cite{zhang2024frnet}. Table~\ref{tab:bench_multidataset} provides a separate 96-step snapshot for MsgNet, AdaWaveNet, TSLANet, and WaveForM across five datasets where source values are available. Because this table contains only the 96-step horizon and has missing entries, it does not support conclusions about long-horizon rankings or general superiority of Fourier- versus wavelet-centric designs.

\begin{table*}[!ht]
    \vspace{-0.05in}
    \centering
    \caption{Benchmarks of Existing Frequency-Based Methods and Comparison}
    \vspace{-0.1in}
    \label{tab:bench_comparision}
    \setlength{\tabcolsep}{15pt}
    \resizebox{0.95\textwidth}{!}{
                \begin{tabular}{c|c|c|c|c|c|c|c}
        \toprule[1.2pt]
        \multirow{2}{*}{\textbf{Methods}} & \multirow{2}{*}{\textbf{Source}} & \multirow{2}{*}{\textbf{\makecell[c]{Frequency \\ Domain}}} & \multirow{2}{*}{\textbf{Metrics}} & \multicolumn{4}{c}{\textbf{Weather}} \\
        \cmidrule(lr){5-8}
         &  &  &  & \textbf{96} & \textbf{192} & \textbf{336} & \textbf{720} \\
        \midrule[1.2pt]
        WaveForM~\cite{yang2023waveform} & AAAI 2023 & Wavelet & \makecell{MSE\\MAE}  & \makecell{0.160\\0.227} & \makecell{0.209\\0.280} & \makecell{0.278\\0.332} & \makecell{0.364\\0.399} \\ \midrule
        WaveRoRA~\cite{liang2024waverora} & Arxiv 2024 & Wavelet & \makecell{MSE\\MAE}  & \makecell{0.159\\0.204} & \makecell{0.208\\0.250} & \makecell{0.263\\0.292} & \makecell{0.345\\0.346}  \\ \midrule
        WFTNet~\cite{liu2024wftnet} & ICASSP 2024 & Wavelet & \makecell{MSE\\MAE}  & \makecell{0.161\\0.210} & \makecell{0.211\\0.254} & \makecell{0.271\\0.296} & \makecell{0.347\\0.346}  \\ \midrule
        Msgnet~\cite{cai2024msgnet} & AAAI 2024 & Fourier & \makecell{MSE\\MAE}  & \makecell{0.163\\0.212} & \makecell{0.212\\0.254} & \makecell{0.272\\0.299} & \makecell{0.350\\0.348}  \\ \midrule
        TSLANet~\cite{eldele2024tslanet} & ICML 2024 & Fourier & \makecell{MSE\\MAE}  & \makecell{0.148\\0.197} & \makecell{0.193\\0.241} & \makecell{0.245\\0.282} & \makecell{0.325\\0.337}  \\ \midrule
        FRNet~\cite{zhang2024frnet} & KDD 2024 & Fourier & \makecell{MSE\\MAE}  & \makecell{\textbf{0.143}\\\textbf{0.195}} & \makecell{\textbf{0.184}\\\textbf{0.235}} & \makecell{\textbf{0.235}\\\textbf{0.275}} & \makecell{\textbf{0.306}\\\textbf{0.330}}  \\ \midrule
        FWin~\cite{tran2023fourier} & Arxiv 2023 & Fourier & \makecell{MSE\\MAE}  & \makecell{-\\-} & \makecell{-\\-} & \makecell{0.338\\0.458} & \makecell{0.291\\0.421}  \\ \midrule
        FEDAF~\cite{yang2024fedaf} & \makecell[c]{Neural Computing and\\ Applications 2024} & Fourier & \makecell{MSE\\MAE}  & \makecell{0.245\\0.329} & \makecell{0.285\\0.342} & \makecell{0.336\\0.371} & \makecell{0.391\\0.403}  \\ \midrule
        AdaWaveNet~\cite{yu2024adawavenet} & Arxiv 2024 & Wavelet & \makecell{MSE\\MAE}  & \makecell{0.169\\0.215} & \makecell{0.203\\0.245} & \makecell{0.248\\0.286} & \makecell{0.313\\0.336}  \\ \midrule
        FITS~\cite{xu2024fits} & ICLR 2024 & Wavelet & \makecell{MSE\\MAE}  & \makecell{\textbf{0.143}\\-} & \makecell{0.186\\-} & \makecell{0.236\\-} & \makecell{0.307\\-}  \\ \midrule
        DLinear +FAN~\cite{ye2024frequency} & Arxiv 2024 & Fourier & \makecell{MSE\\MAE}  & \makecell{0.173\\0.214} & \makecell{0.210\\0.254} & \makecell{0.275\\0.298} & \makecell{0.340\\0.345}  \\ \midrule
        Fredformer~\cite{piao2024fredformer} & KDD 2024 & Fourier & \makecell{MSE\\MAE}  & \makecell{0.163\\0.207} & \makecell{0.211\\0.251} & \makecell{0.267\\0.292} & \makecell{0.343\\0.341}  \\ \midrule
        FreEformer~\cite{yue2025freeformer} & Arxiv 2025 & Fourier & \makecell{MSE\\MAE}  & \makecell{0.153\\\textbf{0.189}} & \makecell{0.201\\0.236} & \makecell{0.261\\0.282} & \makecell{0.341\\0.334}  \\ \midrule
        RecFNO~\cite{zhao2024recfno} & CISAT 2024 & Fourier & \makecell{MSE\\MAE}  & \makecell{0.166\\-} & \makecell{0.210\\-} & \makecell{0.253\\-} & \makecell{0.341\\-}  \\ \midrule
        ATFNet~\cite{ye2024atfnet} & Arxiv 2024 & Fourier & \makecell{MSE\\MAE}  & \makecell{0.156\\0.206} & \makecell{0.199\\0.246} & \makecell{0.249\\0.286} & \makecell{0.311\\0.335}  \\
        \midrule[1.2pt]
        \end{tabular}}
        \vspace{-0.1in}
\end{table*}

\begin{table}[!ht]
    \centering
    \caption{Descriptive cross-dataset snapshot (MSE at the 96-step horizon, as reported in original publications). Bold identifies the smallest available value among the four displayed rows, not a controlled ranking.}
    \vspace{-0.15in}
    \label{tab:bench_multidataset}
    \scriptsize
    \setlength{\tabcolsep}{5mm}{
    \begin{tabular}{c|c|c|c|c|c}
    \toprule
    \textbf{Method} & \textbf{Weather} & \textbf{ETTm1} & \textbf{Electricity} & \textbf{Traffic} & \textbf{Exchange} \\
    \midrule
    Msgnet~\cite{cai2024msgnet} (Fourier) & 0.163 & 0.319 & 0.165 & - & 0.102 \\
    AdaWaveNet ~\cite{yu2024adawavenet} (Wavelet) & 0.169 & 0.326 & 0.146 & 0.417 & 0.086 \\
    TSLANet~\cite{eldele2024tslanet} (Fourier) & \textbf{0.148} & \textbf{0.289} & \textbf{0.136} & \textbf{0.372} & \textbf{0.083} \\
    WaveForM~\cite{yang2023waveform} (Wavelet) & 0.160 & - & 0.153 & 0.567 & - \\
    \bottomrule
    \end{tabular}}\vspace{-0.1in}
    \footnotetext{Values taken from respective source publications; cross-paper comparisons inherit the limitations discussed in Section~\ref{sec:benchmark}.}
\end{table}

\subsection{Cross-Task Comparisons}

Beyond forecasting, spectral methods have been evaluated on classification and anomaly detection. Table~\ref{tab:bench_crosstask} reports exact values from Tables~2 and~4 of the TSLANet paper~\cite{eldele2024tslanet}: mean accuracy over 26 UEA datasets and F1 on SMD. The comparison is source-internal (TSLANet and baselines evaluated within that paper), which is more interpretable than combining values from unrelated protocols. WEFormer~\cite{he2025weformer} provides additional application evidence on WESAD, MOCAS, and two EEG datasets, but we do not merge its values into this table because its physiological small-sample protocols differ. These results provide evidence that frequency-based models are used beyond forecasting; metrics remain task-specific and cannot be compared numerically with forecasting MSE/MAE.

\begin{table}[!ht]
    \centering
    \caption{Cross-task results reported by TSLANet~\cite{eldele2024tslanet}. Accuracy and F1 are percentages; each row is compared only with baselines evaluated under the corresponding source-paper protocol.}
    \vspace{-0.1in}
    \label{tab:bench_crosstask}
    \scriptsize
    \setlength{\tabcolsep}{15pt}
    \begin{tabular}{c|c|c|c|c}
    \toprule
    \textbf{Method} & \textbf{Task} & \textbf{Dataset} & \textbf{Metric} & \textbf{Reported value} \\
    \midrule
    TSLANet~\cite{eldele2024tslanet} & Classification & UEA (26 datasets) & Mean accuracy & 72.73 \\
    Best listed baseline in source (TS-TCC) & Classification & UEA (26 datasets) & Mean accuracy & 69.38 \\
    TSLANet~\cite{eldele2024tslanet} & Anomaly detection & SMD & F1 & 87.91 \\
    Best listed baseline in source (GPT4TS) & Anomaly detection & SMD & F1 & 86.89 \\
    \bottomrule
    \end{tabular}
    \vspace{-0.1in}
\end{table}

\section{Application-Specific Advances} \label{sec:application}

\subsection{Time-Series Forecasting}

Time series forecasting has attracted immense interest due to its wide range of applications, such as in economics~\cite{granger2014forecasting,siami2018forecasting}, healthcare~\cite{alassafi2022time}, energy consumption~\cite{martin2010prediction,qian2019review}, traffic flow~\cite{chen2001freeway,yin2021deep}, and weather forecasting~\cite{karevan2020transductive,wu2023interpretable}. It aims to predict future temporal variations based on historical observations of time series data~\cite{wu2023timesnet}. However, due to the complex and non-stationary nature of real-world systems, the observed series often exhibit intricate temporal patterns, where various types of variations, such as increasing, decreasing, and fluctuating trends, are intertwined, posing significant challenges to the forecasting task~\cite{wang2024timemixer}. Predicting long-term time series trends is a significant challenge, where frequency domain methods excel at capturing long-range patterns. While deep learning models like CNNs~\cite{wu2023timesnet} and Transformers~\cite{nie2023time} have achieved promising results, many leverage the Fourier transform's core property: that convolution in the time domain is equivalent to multiplication in the frequency domain. This principle has spurred the development of efficient convolution-based forecasting models~\cite{zhang2024frnet,cai2024msgnet}.

Early innovations in this area focused on computational efficiency. For example, Park et al. proposed Partial Fourier Transformation (PFT) to calculate a subset of Fourier coefficients more efficiently~\cite{park2021fast}. Other work developed non-periodic Fourier-like algorithms suitable for irregularly sampled data~\cite{lange2021fourier}. More recently, the trend has shifted towards adaptive frequency methods. A prime example is TSLANet~\cite{eldele2024tslanet}, which uses an adaptive spectral block to analyze both short- and long-term dependencies and employs adaptive thresholding to reduce noise, showcasing a more dynamic approach.

\subsubsection{Frequency-enhanced Transformer}

Transformer-based methods for time series forecasting~\cite{zhou2022fedformer,tran2023fourier,ma2023long,zhou2024fourier,yang2024fedaf} aim to provide better performance in the long-term prediction. Many studies have been done to further improve the efficiency~\cite{kitaev2020reformer,zhou2021informer} and effectiveness~\cite{wu2021autoformer,zhou2022fedformer,shabani2023scaleformer} of the Transformer for time series data. Specifically, to capture global-view dependencies of time series, Zhou \textit{et al.}~\cite{zhou2022fedformer} proposed a method, namely FEDformer, to decompose Transformer with Fourier transform to compact representations of long-term time series patterns into frequency domain. Another work~\cite{sasal2022w} introduces an innovative approach to learning representations of univariate time series, named W-Transformer, which is built upon a transformer encoder structure utilizing wavelets. The W-Transformers apply a maximal overlap discrete wavelet transformation to the time series information. Meanwhile, local transformers are adopted to effectively capture the nonstationary nature and intricate long-term nonlinear relationships within the time series data. Different from other studies, Jin \textit{et al.}~\cite{jin2022time} developed a novel approach for generating token sequences tailored for 1D data, namely TST, a fusion of the time series tokenizer and Transformer architecture. More specifically, TST introduces a way to generate token sequences from one-dimensional data, including time series data. This time series tokenizer is then integrated into a Transformer architecture. In this way, good performance is achieved. To capture temporal-spectral correlations effectively~\cite{yang2024graformer,zhang2023self,wang2024card}, to be specific, Zhang \textit{et al.}~\cite{zhang2023self} proposed cross reconstruction transformer (CRT). CRT facilitates time series representation learning by employing a cross-domain dropping-reconstruction task via extracting the frequency domain of the time series using the fast Fourier transform and randomly eliminating specific patches in both the time and frequency domains. Woo \textit{et al.}~\cite{woo2022etsformer} proposed ETSFormer, a fresh Transformer architecture tailored for time-series data. This model leverages the concept of exponential smoothing to enhance Transformers for time-series forecasting. Drawing inspiration from classical exponential smoothing techniques in time-series prediction, ETSFormer introduces the innovative concepts of Exponential Smoothing Attention and Frequency Attention. These mechanisms replace the conventional self-attention module in standard Transformers.

In recent years, frequency-domain-based models have been proposed and have achieved great performance~\cite{yi2023survey}, benefiting from the robust frequency domain modeling. Frequency spectra exhibit strong consistency across different spans of the same series, making them suitable for forecasting. The FEDformer~\cite{zhou2022fedformer} mentioned above performs DFT and sampling prior to Transformer. Fan \textit{et al.}~\cite{fan2022depts} introduce a deep expansion learning framework, DEPTS, for periodic time series forecasting. They use the DFT to capture periodic patterns for better forecasts. Zhou \textit{et al.}~\cite{zhou2022film} designed a Frequency improved Legendre Memory model, named FiLM. FiLM applies Legendre polynomial projections to approximate historical information, uses Fourier projection to remove noise, and adds a low-rank approximation to speed up computation. Yi \textit{et al.} ~\cite{yi2023fouriergnn} transfers GNN operations from the time domain to the frequency domain, and proposed a novel architecture Fourier Graph Neural Network (FourierGNN). FourierGNN accommodates adequate expressiveness and achieves much lower complexity, which can effectively and efficiently accomplish the forecasting. Xu \textit{et al.}~\cite{xu2024fits} applied a low-pass filter and complex-valued linear projection in the frequency domain. To address the problem of MLP-based forecasting methods suffering from the point-wise mappings and information bottleneck, Yi \textit{et al.}~\cite{yi2023frequency} explore a novel direction of applying MLPs in the frequency domain for time series forecasting. Their model, FreTS, employs frequency-domain MLPs to model channel and temporal dependencies. Similarly, Fan \textit{et al.}~\cite{fan2024deep} introduced a
Fourier derivative operator to address non-stationarity, and Ye \textit{et al.}~\cite{ye2024frequency} introduces frequency adaptive normalization for non-stationary data.

These existing frequency-domain-based works rely on linear layers to learn frequency-domain representations, resulting in a performance gap. Recently, Fredformer~\cite{piao2024fredformer} applies the vanilla Transformer to patched frequency tokens to address the frequency bias is- sue. However, the patching technique introduces additional hyper-parameters and undermines the inherent global perspective~\cite{yi2023frequency} of frequency-domain modeling. To address these issues, Yue \textit{et al.}~\cite{yue2025freeformer} applied the Transformer to frequency-domain variate tokens for representation learning. Specifically, they embed the entire frequency spectrum as variate tokens and capture cross-variate dependencies among them. This architecture offers three main advantages: 1) simple frequency-domain operations can correspond to complex temporal operations; 2) Inter-variate correlations typically exists and learning these correlations could be beneficial for more robust frequency representation; 3) The permutation- invariant nature of the attention mechanism naturally aligns with the order-insensitivity of variates. Unlike previous models in the time domain, Yue \textit{et al.}'s approach~\cite{yue2025freeformer} shifts the focus to the frequency domain to explore dependencies among the frequency spectra of multiple variables for more robust representations.

\subsubsection{Multi-scale Decomposition Methods} Although Transformer-based methods excel at modeling long-range dependencies due to the self-attention mechanisms~\cite{vaswani2017attention}, they come with computational complexity that scales quadratically with the length of the sequence. Despite their recognized performance, these methods still face challenges in industrial applications, including low training efficiency and high memory consumption. Additionally, self-attention can diminish the temporal relationships when extracting semantic correlations between pairs in long sequences~\cite{zeng2023transformers}, leading to an overemphasis on mutation points and resulting in overfitting. In contrast, MLP-based methods boast significantly lower computational complexity compared to Transformer-based methods. Moreover, MLP-based methods can chronologically model the temporal dynamics in consecutive points, which is crucial for time series analysis~\cite{zeng2023transformers,das2023long}. However, the simplicity of linear mappings in existing MLP-based methods presents an information bottleneck~\cite{liu2024kan}, hindering their ability to capture diverse temporal patterns and limiting their predictive accuracy~\cite{ni2024mixture}.

Motivated by the above observations, Wang \textit{et al.}~\cite{wang2024timemixer} proposed TimeMixer as a fully MLP-based architecture with past-decomposable-mixing and future-multipredictor-mixing blocks~\cite{tolstikhin2021mlp,lee2022fnet} to take full advantage of disentangled multiscale series~\cite{liu2021pyraformer,liu2022scinet} in both past extraction and future prediction phases. Furthermore, Hu \textit{et al.}~\cite{hu2025adaptive} decomposed the time series at multiple scales to precisely discern the intertwined temporal patterns within the complex series, rather than merely breaking it down into seasonal and trend components~\cite{wu2021autoformer,zeng2023transformers}. Unlike the average aggregation of TimeMixer~\cite{wang2024timemixer}, to account for the varying impacts of different temporal patterns on the future, Hu \textit{et al.}~\cite{hu2025adaptive} employed an autocorrelation approach to model their contributions and adaptively integrate these multi-scale temporal patterns based on their respective influences.

Additionally, regarding the special composition and the complex temporal patterns of time series data, early approaches employ manually designed rules or function models to decompose the time series data, such that the temporal patterns can be disentangled and modeled separately~\cite{theodosiou2011forecasting,dagum2016seasonal,wen2019robuststl,dokumentov2022str}. The decomposition usually consists of several components representing different temporal patterns, plus a residual which is supposed to be noise with no useful information. In recent years, deep learning has been widely applied in time series analysis for its strong expressiveness and scalability on large and complex datasets. By combining deep learning with decomposition, many studies show satisfactory results in time series forecasting, such as N-BEATS~\cite{oreshkinn2020n-beats}, N-HiTS~\cite{challu2023nhits} and ETSformer~\cite{woo2022etsformer}. However, N-BEATS~\cite{oreshkinn2020n-beats} and N-HiTS~\cite{challu2023nhits} do not consider the inter-channel correlation, which has been shown critical in multivariate time series analysis tasks. In addition, they are based on plain MLP on the temporal dimension while ETSformer~\cite{woo2022etsformer} is based on self-attention for temporal modeling, all of which do not take into account the sub-series level features. Furthermore, they simply ignore the residual of the decomposition, which may lead to incomplete decomposition that meaningful temporal patterns can be left in the residual and not utilized. Besides, all these schemes have only been tested on the forecasting task, leaving other analysis tasks such as imputation, anomaly detection, and classification unexplored. In contrast, Zhong \textit{et al.}~\cite{zhong2024multi} proposed MSD-Mixer that advances them with multi-scale temporal patching and multi-dimensional MLP mixing for multi-scale sub-series and inter-channel modeling. MSD-Mixer is based exclusively on MLPs, which is simple but effective for time series modeling. MSD-Mixer explicitly decomposes the input time series into different components by generating their latent representations in different layers, and accomplishes the analysis task based on such representations. This idea inspired later studies~\cite{zhang2025multiscale}.

\subsection{Time-Series Anomaly Detection}
In time-series anomaly detection, point-level class imbalance and spectral visibility are related but distinct issues. Point anomalies are typically rare relative to normal observations, which creates a class-imbalance problem for supervised learners; a frequency transform does not, by itself, correct that imbalance. Its separate benefit is representational: filtering can suppress irrelevant high-frequency noise, while sustained or slowly evolving anomalies may produce persistent spectral deviations that are easier to quantify than in the raw time domain, depending on the analysis window and baseline non-stationarity. Long-duration anomalies may be less severely imbalanced than isolated point anomalies, so improved spectral visibility should not be interpreted as a remedy for class rarity.

Time-series motifs provide a related application mechanism. A motif is a recurring subsequence whose repeated shape may have domain meaning~\cite{lin2007motif}; motif frequency, duration, and spectral signature can support classification, monitoring, and anomaly detection when a normally recurring pattern disappears or changes. Motif discovery is therefore complementary to global spectral summaries and is discussed further from an interpretability perspective in Section~\ref{sec:challenge_details}.
\subsubsection{Spectral Signature Analysis}
Spectral signature analysis~\cite{elsner2013singular,shi2025characterizing,wang2025time,vinogradov2025signatures,toth2024learning} has emerged as a powerful framework for time series anomaly detection, leveraging frequency-domain representations to identify deviations from normal behavior. Traditional anomaly detection methods often rely on statistical thresholds or deep learning reconstruction errors, but spectral approaches offer unique advantages by decomposing signals into their frequency components and analyzing their energy distributions. Recent studies have demonstrated significant improvements in detection accuracy, robustness to noise, and interpretability using spectral techniques, particularly in applications ranging from industrial monitoring to healthcare diagnostics. A recent study~\cite{wang2025time} proposes a method, namely MultiPSCA, centering on exploring path signatures - a powerful mathematical transformation that inherently possesses time warping invariance properties. MultiPSCA investigates innovative augmentation techniques for time series data that enable their signature space representations to effectively capture and reflect typical warping constraints. Through rigorous analysis, MultiPSCA establishes that comparing path signatures fundamentally corresponds to evaluating time series using elastic distance metrics.  MultiPSCA bridges the gap between theoretical signature properties and practical implementation considerations for real-world time series analysis. During necking deformation, acoustic emission activity tends to decline as strain hardening diminishes and dislocation motion becomes restricted in the final deformation stages. To interpret the acoustic emission spectral characteristics observed during localized plastic flow, Vinogradov et al.~\citet{vinogradov2025signatures} apply a phenomenological model grounded in dislocation dynamics - originally developed for uniform deformation scenarios. The findings highlight the acoustic emission as a promising tool for non-destructive assessment and early failure detection in structural metals, demonstrating its responsiveness to microstructural evolution and plastic instabilities.

Former studies~\cite{batista2022optical,gregnanin2023signature,bullock2020monitoring,zulfa2021spectral} based on spectral signature analysis for time series are shown as follows:~\citet{batista2022optical} investigates the performance of multiple machine learning models when applied to three distinct datasets: (1) a single composite image, (2) a time-series composite of the same location, and (3) a preprocessed version of the time-series composite using basic temporal analysis methods. Results confirm the anticipated outcome - classification models trained on single composite imagery demonstrate notably lower median producer's accuracy and user's accuracy for individual classes compared to their overall accuracy across all classes. This performance gap highlights the inherent limitations of static composite imagery for detailed classification tasks compared to time-series-based approaches. Gregnanin et al.~\citet{gregnanin2023signature} introduce a novel framework utilizing the signature matrix to analyze complex time series data. This signature-based approach provides a robust similarity metric that addresses several limitations inherent in conventional correlation analyses.

\subsubsection{Wavelet-based Change Detection}
Wavelet-based methods~\cite{ghaderpour2021survey} have emerged as a powerful tool for detecting changes and anomalies in time series data by leveraging their ability to localize both frequency and temporal information. Unlike traditional Fourier analysis, which provides only global frequency content, wavelet transforms decompose signals into multi-resolution components, capturing abrupt changes, transient events, and gradual trends simultaneously. By applying discrete or continuous wavelet transforms (DWT/CWT), key features such as energy distribution, scale-wise variations, and localized discontinuities can be extracted. Statistical thresholding, machine learning classifiers, or entropy-based measures are then applied to wavelet coefficients to identify significant deviations from normal behavior. This approach is particularly effective in applications such as structural health monitoring (detecting cracks or vibrations), environmental sensing (identifying pollution spikes), and financial market analysis.

Recent advances~\cite{zhang2025stwanet,fonseca2024time,wang2024w,yu2024adawavenet} integrate deep learning with wavelet decomposition, enhancing detection accuracy in noisy or non-stationary data. Compared to raw time-domain methods, wavelet-based change detection offers superior sensitivity to subtle perturbations while maintaining interpretability through its multi-scale framework. Existing building change detection methods frequently face challenges in accurately identifying complex architectural boundaries and subtle surface variations. To overcome these limitations, ~\citet{zhang2025stwanet} present STWANet - an innovative deep learning framework that synergistically combines multi-scale feature extraction with frequency-domain analysis. The architecture begins with a ResNet18 backbone to generate rich hierarchical features containing both detailed spatial information and high-level semantics. At its core, STWANet employs three specialized components working in concert: (1) a differential self-attention mechanism that automatically identifies and weights significant spatio-temporal variations across feature scales, (2) a wavelet-based enhancement module that specifically amplifies high-frequency components critical for edge preservation, and (3) a parallel dual-attention pathway that maintains comprehensive spatial context while focusing on change regions. This multi-branch design ensures robust handling of diverse building characteristics, from intricate facade details to large-scale structural elements. The wavelet transform component plays a particularly vital role by operating in the frequency domain to isolate and enhance fine-grained texture patterns that often indicate meaningful changes. Meanwhile, the complementary attention mechanisms work to preserve both local precision and global consistency throughout the analysis process. Together, these innovations address the common shortcomings of existing approaches, delivering superior performance in capturing the complete spectrum of building modifications while maintaining accuracy for complex urban geometries. ~\citet{fonseca2024time} present an unsupervised approach for identifying spatio-temporal variations in multi-temporal SAR imagery, termed Time-Referenced Wavelet Energy Correlation Screening (TR-WECS). TR-WECS operates on sequential SAR acquisitions of a target area, utilizing the initial image as a temporal benchmark for change assessment. The technique incorporates two key processing stages: first, wavelet-based smoothing is applied to mitigate speckle-induced artifacts, ensuring detected variations represent genuine surface modifications rather than noise. Subsequently, the analysis examines temporal patterns across non-contiguous image blocks of varying dimensions through innovative wavelet energy distribution analysis combined with high-dimensional correlation assessment. This dual-phase approach enables robust change detection while maintaining sensitivity to both subtle and significant alterations across the temporal sequence.

Earlier studies~\cite{chaovalit2011discrete,grane2010wavelet,alarcon2009change,bilen2002wavelet,li2021w} based on wavelet transform are shown as follows: Financial time series~\cite{grane2010wavelet} often contain extreme values that can distort parameter estimation, produce misleading statistical conclusions, and degrade forecasting performance, particularly in volatility modeling. Addressing these anomalous observations requires careful consideration in econometric analysis. ~\citet{grane2010wavelet} introduce a wavelet-based framework for identifying and adjusting outliers that is compatible with numerous volatility specifications.

\subsection{Time-Series Classification}
While anomaly detection focuses on identifying rare deviations from normal patterns, time-series classification addresses the broader challenge of categorizing entire sequences into predefined classes. This task faces unique challenges due to complex temporal patterns where subtle variations in timing, frequency, and amplitude distinguish between categories. The computational complexity of traditional approaches, which require exhaustive searches across candidate subsequences, further complicates the process. Additionally, cross-domain variations make it difficult for models to capture distinguishing features while maintaining generalization. Frequency domain analysis proves particularly valuable by decomposing complex temporal patterns into constituent frequency components, revealing underlying periodic structures and energy distributions crucial for accurate classification. This transformation provides a robust foundation for distinguishing between different time-series categories while enabling efficient feature extraction and cross-domain knowledge transfer.

\subsubsection{Discriminative Frequency Features}
Time series classification can be significantly enhanced by leveraging discriminative frequency features, which capture unique spectral patterns indicative of different classes. Unlike traditional time-domain approaches, frequency-based methods decompose signals using techniques such as Fourier transforms, wavelet analysis, or power spectral density estimation, extracting meaningful oscillations, harmonics, and energy distributions that distinguish between categories. These features are particularly effective for applications where periodic patterns, transient events, or noise robustness are critical, such as in ECG arrhythmia detection, industrial fault diagnosis, or financial trend analysis. Advanced methods~\cite{he2025weformer,farahani2025time,ahamed2025tscmamba,liu2025financial,huang2025revisiting,gackstetter2025self,fan2025towards} combine these spectral representations with machine learning models, including convolutional neural networks (CNNs) or attention mechanisms, to improve classification accuracy. Recent innovations also explore adaptive frequency band selection and multiresolution analysis to optimize discriminative power while reducing computational overhead. ~\citet{he2025weformer} present WEFormer - a novel classification framework specifically designed for small-sample learning scenarios. WEFormer leverages a pre-trained time series foundation model with fixed parameters as a feature encoder, eliminating manual feature engineering while preserving robust representational capacity.

Many existing shapelet-based approaches face high computational costs due to the exhaustive search required across numerous candidate subsequences. To address this challenge, ~\citet{zhang2018discriminative} introduce an efficient shapelet discovery technique that treats time series as high-dimensional vectors and identifies the most discriminative dimensions representing optimal shapelet locations. ~\citet{zhang2018discriminative} integrate Local Fisher Discriminant Analysis (LFDA) with dual sparsity constraints designed to preserve the temporal continuity inherent in time series data. This combined approach not only reduces computational overhead but also enhances the interpretability of extracted shapelets by emphasizing continuous, class-relevant patterns.

\subsubsection{Cross-domain Representations}
Effective time series classification often requires robust feature representations that generalize across diverse domains. Cross-domain representation learning addresses this challenge by extracting transferable patterns that capture underlying temporal dynamics while minimizing domain-specific biases. Recent approaches leverage deep neural networks to learn domain-invariant features through techniques such as adversarial training, maximum mean discrepancy (MMD) minimization, or contrastive learning in shared latent spaces. These methods enable knowledge transfer between source and target domains with different statistical distributions, particularly valuable in scenarios with limited labeled data. By transforming raw time series into meaningful representations that preserve discriminative characteristics across domains, these techniques~\cite{cheng2025cross,jia2025contrastive,huang2021artificial,liu2024timemmd,hu2025fusion,lan2025cicada,peng2024cross} improve classification performance in applications ranging from industrial sensor monitoring to healthcare diagnostics. To be specific, Cheng et al.~\citet{cheng2025cross} propose a method, namely CrossTimeNet, an innovative self-supervised learning framework for cross-domain temporal representation learning. The architecture tackles data scarcity challenges through several key innovations: First, it employs a transformer-based language model as its foundational encoder, leveraging its inherent capability to model complex sequential dependencies. Second, CrossTimeNet implements a masked reconstruction objective that explicitly preserves the local temporal structure of the input while learning generalizable patterns. To bridge the gap between continuous time-series data and discrete sequence modeling, CrossTimeNet develops a specialized tokenization layer incorporating vector quantization to enable effective discrete representation of continuous waveforms. CrossTimeNet enables domain-agnostic pretraining while maintaining sensitivity to temporal locality, ultimately enhancing performance on downstream tasks.

\section{Challenges and Limitations of Frequency Methods}
\label{sec:challenge_details}

\begin{wrapfigure}{r}{0.55\textwidth}
    \vspace{-0.2in}
    \centering
    \includegraphics[width=0.5\textwidth]{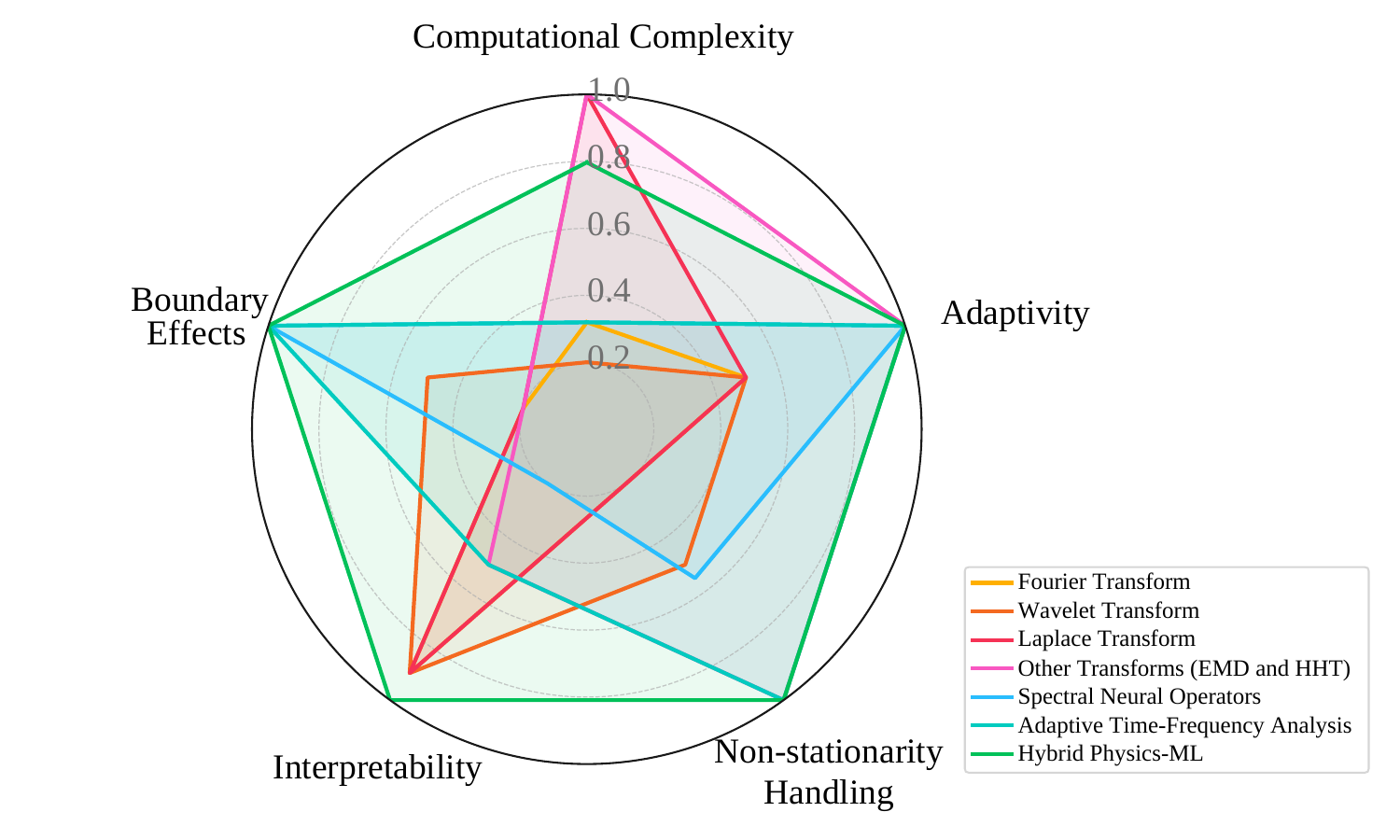}
    \setlength{\abovecaptionskip}{0.5mm}
    \caption{Qualitative comparison of classical and modern frequency methods. Every plotted score is oriented so that larger is preferable: the axis displayed as ``Computational Complexity'' should be read as computational efficiency (lower cost receives a larger score), and ``Boundary Effects'' should be read as robustness to boundary effects (fewer artifacts receives a larger score). Scores use the coarse rubric described in the Table~\ref{tab:comparison_table_challenges} footnote and are not measured benchmark values.}
    \label{fig:frequency_methods_radar}
    \vspace{-0.1in}
\end{wrapfigure}

The integration of classical and modern frequency methods in time series analysis presents a range of technical challenges across computational and interpretability dimensions. In this section, we systematically examine five key performance indicators: computational complexity, non-stationarity handling, interpretability versus performance trade-offs, boundary and finite-length issues, and scalability to high-dimensional time series. For each indicator, we compare classical and modern approaches, analyze their respective advantages and limitations, and highlight potential directions for integration. To facilitate a holistic understanding, we include both a comparative table (Table~\ref{tab:comparison_table_challenges}) and graphical radar chart (Figure~\ref{fig:frequency_methods_radar}) that visually summarize the relative strengths of each method.

\subsection{Computational Complexity}

In frequency-domain approaches, computational complexity is a very important metric. Classical frequency-domain methods vary significantly in their computational complexities. The Fourier Transform, especially the Fast Fourier Transform (FFT), is renowned for its computational efficiency, operating at $\mathcal{O}(N \log N)$ complexity. Wavelet Transforms also exhibit relatively moderate complexity, typically scaling as $\mathcal{O}(N)$ to $\mathcal{O}(N^2)$, depending on the chosen wavelet and implementation specifics. Laplace Transforms, while analytically powerful, generally require numerical approximations (such as numerical inversion techniques) in practical applications, making their complexity highly dependent on implementation details and often leading to increased computational overhead. Empirical Mode Decomposition (EMD) and Hilbert-Huang Transform (HHT), although highly effective for non-linear and non-stationary data, exhibit higher complexity due to iterative sifting procedures, typically at least $\mathcal{O}(N^2)$, which poses challenges for large-scale datasets.

On the other hand, modern frequency analysis approaches typically demonstrate high computational complexity due to intensive training and inference requirements. Spectral Neural Operators utilize deep neural network structures, whose complexity scales with model depth, network architecture, and the dimensionality of data, often resulting in computational loads significantly higher than classical approaches, particularly during training phases. Adaptive Time-Frequency Analysis methods integrate classical transforms with adaptive or data-driven modules, typically incurring increased computation due to the optimization steps inherent in adaptivity. Hybrid Physics-ML approaches combine mechanistic models and machine learning techniques, leading to complexities that vary widely depending on the proportion of physics-based modeling versus machine learning components, generally incurring substantial computational cost from iterative parameter estimation and model integration processes.

\begin{table*}[!ht]
    \vspace{-0.1in}
    \renewcommand\arraystretch{3.5}
    \centering
    \caption{Challenges and limitations of classical methods and modern methods. \colorbox{green!30}{green}, \colorbox{orange!30}{orange}, and \colorbox{red!30}{red} indicate excellent or high, moderate or acceptable, and poor or low respectively.}
        \vspace{-0.15in}
    \label{tab:comparison_table_challenges}
        \scriptsize
    \setlength{\tabcolsep}{10pt}
    \resizebox{\textwidth}{!}{
                \begin{tabular}{c|c|c|c|c|c|c}
        \toprule[1.2pt]
        \multicolumn{2}{c|}{\textbf{Frequency Approaches}} & \makecell[c]{\textbf{Computational Complexity}\footnotemark} & \textbf{Adaptivity} & \makecell[c]{\textbf{Non-Stationary Handling}} & \textbf{\makecell[c]{Interpretability}} & \textbf{\makecell[c]{Boundary Effects}}  \\
        \midrule[1.2pt]
        \multirow{5}{*}{\makecell[c]{\textbf{Classical} \\ \\ \textbf{Methods}}} & Fourier Transform & \colorbox{green!30}{$\mathcal{O}(N \log N)$-$\mathcal{O}(N^2)$} & \colorbox{orange!30}{Fixed Basis} & \colorbox{red!30}{Poor} & \colorbox{green!30}{High} & \colorbox{red!30}{Severe}\\
        \cmidrule(r){2-7}
        ~  & Wavelet Transform & \colorbox{green!30}{$\mathcal{O}(N)$-$\mathcal{O}(N^2)$} & \colorbox{orange!30}{Fixed Basis} & \colorbox{orange!30}{Moderate} & \colorbox{green!30}{High} & \colorbox{orange!30}{Moderate}\\
        \cmidrule(r){2-7}
        ~  & Laplace Transform & \colorbox{red!30}{$\mathcal{O}(N^2)$} & \colorbox{orange!30}{Fixed Basis} & \colorbox{red!30}{Poor} & \colorbox{green!30}{High} & N/A\\
        \cmidrule(r){2-7}
        ~  & \makecell[c]{Other Transforms \\ (\textit{e.g.,} EMD and HHT)} & \colorbox{red!30}{$\mathcal{O}(N^2)$-$\mathcal{O}(N^3)$} & \colorbox{green!30}{Data-adaptive} & \colorbox{green!30}{Excellent} & \colorbox{orange!30}{Moderate} & \colorbox{red!30}{Severe}\\
        \cmidrule(r){1-7}

        \multirow{3}{*}{\makecell[c]{\textbf{Modern} \\ \\ \textbf{Methods}}} & \makecell[c]{Spectral Neural \\ Operators} & \colorbox{green!30}{$\mathcal{O}(N \log N)$} & \colorbox{green!30}{Learned} & \colorbox{orange!30}{Moderate} & \colorbox{red!30}{Poor} & \colorbox{green!30}{Learned}\\
        \cmidrule(r){2-7}
        ~  & \makecell[c]{Adaptive Time-\\Frequency Analysis} & \colorbox{green!30}{$\mathcal{O}(N \log N)$-$\mathcal{O}(N^2)$} & \colorbox{green!30}{Learned} & \colorbox{green!30}{Excellent} & \colorbox{orange!30}{Moderate} & \colorbox{green!30}{Learned}\\
        \cmidrule(r){2-7}
        ~  & Hybrid Physics-ML & \colorbox{orange!30}{$\mathcal{O}(N)$-$\mathcal{O}(N^3)$} & \colorbox{green!30}{Learned + Physics} & \colorbox{green!30}{Excellent} & \colorbox{green!30}{High} & \colorbox{green!30}{Physics-regularized}\\
        \bottomrule[1.2pt]
        \end{tabular}
    }
    \vspace{-0.2in}
\end{table*}
\footnotetext{Complexities are asymptotic costs under the assumptions stated by the cited algorithms, not exact operation counts. A direct $N$-point DFT costs $\mathcal{O}(N^2)$, whereas an FFT implementation costs $\mathcal{O}(N\log N)$; these are different algorithms and should not be interpreted as uncertainty around one cost. Neural-method costs additionally depend on retained modes, channels, layers, training, and hardware. The radar uses three qualitative efficiency tiers: high for approximately linear or $N\log N$ transform cost, moderate for quadratic cost or substantial learned overhead, and low for higher/iterative cost. The chart is a visual summary rather than a numerical benchmark.}

\subsection{Adaptivity}

Adaptivity refers to a method’s ability to tailor its analysis dynamically to the structure of the data, especially in scenarios where signal characteristics vary over time or across dimensions. Classical frequency-domain methods are generally built on fixed basis functions, which limits their flexibility. For example, the Fourier Transform uses global sinusoidal bases, which are efficient but rigid, resulting in poor adaptability to signals with local variations or time-varying spectra. Similarly, both Wavelet and Laplace Transforms rely on predefined basis sets or integral kernels, which, although they offer some resolution across scales, are not learned or data-driven. Consequently, while efficient and interpretable, these classical methods often fall short in modeling localized or highly non-stationary phenomena. Only more advanced classical techniques like Empirical Mode Decomposition (EMD) and Hilbert-Huang Transform (HHT) exhibit data-adaptive characteristics by decomposing signals into components derived directly from intrinsic oscillations, rather than from predefined bases.

Modern methods are inherently more adaptive. Spectral Neural Operators, for instance, can learn spectral bases from data, adjusting their frequency representations during training to suit complex or irregular time series. Adaptive Time-Frequency Analysis methods represent a hybrid class, often extending wavelet or Fourier-based techniques by introducing learnable or data-conditioned parameters that modify the basis functions or windowing dynamically. Hybrid Physics-ML methods push this further by embedding learned components into physical models, effectively combining domain knowledge with flexible representation learning. This enables them to handle a broad spectrum of signal complexities, from smooth to chaotic, without relying on handcrafted bases.

\subsection{Spectral Artifacts and Sampling Effects}

Practical frequency-domain preprocessing often determines whether spectral methods help or hurt downstream performance. \textit{Irregular or non-uniform sampling} violates the regular-grid assumption of the standard FFT and can bias frequency and phase estimates. Appropriate alternatives include nonuniform transforms, Lomb--Scargle-type estimators, or graph/wavelet extensions~\cite{lange2021fourier,defferrard2016convolutional}, selected according to the sampling process.

\textit{Spectral leakage} arises when a finite segment is implicitly treated as periodic and its endpoints do not match. Windowing and multitaper estimators reduce sidelobes or variance with corresponding resolution trade-offs. Zero-padding does not reduce leakage; it only samples the existing spectrum more densely. \textit{Aliasing} occurs when energy above the post-sampling Nyquist frequency folds into lower frequencies. It should be controlled with an anti-alias filter before downsampling; imputation can add spectral artifacts but is not itself a Nyquist violation.

\textit{Boundary effects} arise because finite-window Fourier and wavelet calculations require an implicit extension outside the observed interval. Periodic, zero, symmetric, or predictive extensions create different edge artifacts, and results near the boundaries should be reported or excluded according to the transform's cone of influence. Finally, learned filters can fit protocol-specific artifacts; robustness therefore requires matched or deliberately varied sampling, windowing, and padding protocols across training and deployment.

\subsection{Non-Stationary Handling}

One of the most pervasive challenges in time series analysis is dealing with non-stationarity—patterns that evolve over time in unpredictable ways. Classical frequency-based approaches exhibit varying capacities in handling non-stationarity. The Fourier Transform, while powerful for analyzing stationary signals, assumes global stationarity and lacks temporal localization, making it unsuitable for signals with time-varying spectral characteristics. The Wavelet Transform addresses this limitation by offering multi-resolution analysis, which allows for localization in both time and frequency domains, thus better capturing non-stationary behavior. Similarly, the Laplace Transform, although not as commonly used in time series analysis, provides some temporal adaptability but remains limited by its reliance on analytic continuation and integral transforms. More advanced classical methods like EMD and the HHT are explicitly designed to handle non-linear and non-stationary signals by decomposing signals into intrinsic mode functions without requiring predefined basis functions, offering significant improvements in adaptivity.

In contrast, modern methods inherently address non-stationarity by design. Spectral Neural Operators can learn non-stationary dynamics through data-driven spectral convolutions, where the model adapts to local patterns and shifts in the frequency domain during training. Adaptive Time-Frequency Analysis techniques further extend classical methods by incorporating data-adaptive parameters and context-specific spectral estimation, enabling dynamic response to changing statistical properties. Hybrid Physics-ML approaches, which combine domain knowledge with machine learning, also enhance non-stationarity handling by embedding physical constraints while allowing data-driven flexibility. These methods often leverage attention mechanisms, recurrent structures, or time-aware embeddings to effectively model temporal variability across multiple scales.

\subsection{Interpretability}

Interpretability remains a key concern in frequency-based time series analysis, particularly for applications that demand transparency, explainability, and trust. Classical methods generally offer strong interpretability due to their well-defined mathematical foundations and human-understandable spectral outputs. Fourier and Wavelet Transforms both produce meaningful frequency-domain representations, with the latter enhancing interpretability by enabling time-frequency localization. Laplace Transforms preserve analytic clarity through transform-domain pole-zero behavior, while EMD and HHT contribute intuitive decomposition into intrinsic mode functions (IMFs), although with somewhat less mathematical formalism.

In contrast, modern methods often trade off interpretability for adaptability and performance. Spectral Neural Operators (SNO), for instance, provide strong modeling capacity but suffer from low interpretability due to their black-box architecture. Adaptive Time-Frequency methods and Hybrid Physics-ML approaches attempt to bridge this gap: the former integrates learned adaptivity while maintaining structural transparency to some extent, and the latter incorporates domain knowledge and physical constraints into learning-based pipelines, achieving high interpretability. Time series \textit{motifs}, recurring subsequences with domain meaning, offer a complementary interpretability axis: motif discovery~\cite{lin2007motif} often corresponds to characteristic spectral signatures, linking time-domain patterns to frequency features for explainable monitoring and classification. These hybrid approaches show that interpretability and learning capacity are not mutually exclusive.

\subsection{Boundary Effects}

Boundary effects represent a persistent challenge in frequency-based time series analysis, especially when working with finite-length or non-periodic signals. Classical methods are particularly vulnerable to edge distortions. The Fourier Transform assumes signal periodicity. Wavelet Transforms handle boundaries more gracefully due to their localized basis functions but still suffer from moderate edge issues, especially for large-scale wavelets. Laplace Transforms, being defined over infinite or semi-infinite domains, are inherently ill-suited for bounded signals, and their applicability to finite-length data remains limited in practice. EMD and HHT are highly adaptive but prone to severe edge effects due to their reliance on extrema and envelope estimation.

Modern methods offer more robust mechanisms to address boundary issues. Spectral Neural Operators and Adaptive Time-Frequency Analysis both integrate learned context padding and local representations that generalize better to edge regions. These are designed to flexibly handle finite-length signals. Most notably, Hybrid Physics-ML methods embed physical boundary constraints into model architecture, enabling improved robustness and generalization at the signal edges.

\section{Future Directions}
\label{sec:future}

\textbf{Limitations of current LLMs.} The directions below envision integrating large language models (LLMs) with frequency-domain time-series analysis. Present-day performance is task-dependent: LLMs exhibit constrained mathematical reasoning, unreliable causal inference, hallucination, and limited cross-domain extrapolation~\cite{mirzadeh2025gsm,kambhampati2024reason}. Frequency-aware tokenization, natural-language spectral explanations, and causal reasoning in Table~\ref{tab:explanation_framework} should therefore be understood as long-term research hypotheses, not established capabilities. Evaluation must separate the numerical encoder from the language component and test calibration, domain shift, causal identifiability, and faithfulness. High-stakes deployment would additionally require symbolic or physics-grounded verification rather than reliance on generated explanations alone.

The integration of frequency-domain techniques with large language models (LLMs) presents testable research opportunities for time-series analysis, subject to the limitations above. Building on the methods compared in Table~\ref{tab:comparison_table_challenges}, we identify four hypotheses for future study; they should not be read as demonstrated breakthroughs.

\subsection{Hybrid Neuro-Symbolic Architectures}
Modern LLMs struggle with the numerical and structural properties of time-series data. We propose two hypotheses rather than established solutions: \textbf{1) Frequency-aware tokenization}: wavelet features could be combined with language-model temporal-context embeddings, but benefit must be tested against numerical-encoder-only controls and under domain shift. \textbf{2) Physics-informed attention}: dominant-frequency estimates could encourage attention among harmonically related components. This bias does not ensure compliance with physical laws; compliance requires explicit constraints and independent residual or conservation tests.

\subsection{Interpretable Frequency-Grounded Explanations}
Current LLMs exhibit critical limitations in explainability for time series tasks, often producing "black box" predictions. Frequency methods offer a principled pathway to address this through mathematically grounded explanations, as systematically compared in Table~\ref{tab:explanation_framework}. The framework establishes direct mappings between spectral features and human-understandable concepts.

\begin{table}[htbp]
    \vspace{-0.05in}
    \centering
    \setlength{\tabcolsep}{15pt}
    \caption{Proposed Frequency-Grounded Explanation Framework for Time Series LLMs}
    \vspace{-0.1in}
    \label{tab:explanation_framework}
    \resizebox{0.85\linewidth}{!}{
    \begin{tabular}{ccc}
    \toprule
    \textbf{Component} & \textbf{Frequency Tool} & \textbf{LLM Integration} \\
    \midrule
    Feature Attribution & Wavelet coherence & Natural language justification \\
    \addlinespace[0.2cm]
    Change Point Detection & HHT instantaneous freq & Causal reasoning (long-term vision) \\
    \addlinespace[0.2cm]
    Anomaly Localization & Spectral residuals & Contrastive explanations \\
    \bottomrule
    \end{tabular}}
    \vspace{-0.1in}
\end{table}

The framework in Table~\ref{tab:explanation_framework} is intended to generate candidate frequency-grounded explanations; their precision, faithfulness, and interpretability require empirical validation. For instance, wavelet coherence values can provide numerical features, but their use as explanation weights requires calibration and comparison with perturbation-based evidence.

\subsection{Few-Shot Learning via Spectral Prototypes}
The data-hungry nature of LLMs poses significant challenges for time series applications where labeled examples are scarce. Frequency-domain representations offer an elegant solution through:

\textbf{1) Spectral memory banks}:
Spectral Memory Banks construct reusable frequency-domain prototypes from unlabeled data, capturing physically meaningful patterns. For new tasks, few-shot learning is achieved by matching spectral signatures of labeled examples to these prototypes, enabling cross-task generalization, physical interpretability, and sample-efficient adaptation with minimal annotations.

\textbf{2) Frequency-based prompting}: This approach constructs more effective prompts by converting time series into frequency-domain features, enhancing language models' understanding of time series. The method extracts key frequency components, converting them into structured prompt information that guides models to generate more accurate time series descriptions, particularly suitable for analyzing data with complex periodic patterns.

\subsection{Cross-Modal Time-Frequency Understanding}
Bridging the semantic gap between numerical time series and textual descriptions requires innovative joint representations: \textbf{1) Joint embedding spaces}: Joint embedding spaces map time series data into the same semantic space as text descriptions through a systematic transformation process. This transformation makes time series embeddings approximately equivalent to LLM text embeddings, enabling direct comparison between signal patterns and textual descriptions, providing a unified framework for cross-modal understanding. \textbf{2) Frequency-conditioned generation}: Frequency-conditioned generation constrains generation through probability decomposition, ensuring outputs respect specific frequency components for physically plausible results. The method translates natural language descriptions into precise mathematical constraints, ensuring generated time series meet text requirements while maintaining physical consistency, providing an effective framework for controllable generation.

\section{Conclusion}
\label{sec:conclu}
This survey confirms frequency-domain analysis as a cornerstone of time series research, while highlighting that its future progress depends on solving key theoretical and practical challenges. By synthesizing over 100 studies, we introduce a unified taxonomy that connects classical Fourier techniques to modern neural operators. This framework, along with proposed benchmarks, offers a robust foundation for comparing the performance of diverse spectral methods. Our review identifies three critical research frontiers: preserving causal structures, quantifying uncertainty in frequency representations, and extending analysis to non-Euclidean topologies. For practitioners, this work provides a systematic guide for method selection. For researchers, it charts a clear path forward, illuminating significant knowledge gaps and pinpointing promising directions at the intersection of spectral methods with geometric deep learning and quantum-enhanced analysis. By defining these challenges, this survey aims to catalyze the next wave of progress in this powerful and dynamic domain.

\bibliographystyle{ACM-Reference-Format}
\bibliography{reference}

\end{document}